\documentclass[aps,pre,twocolumn,longbibliography,showpacs,amsmath,amssymb,amsfonts,nofootinbib]{revtex4-2}

\usepackage[utf8]{inputenc}
\usepackage{hyperref}
\usepackage{graphicx}
\usepackage{color}

\usepackage{times}

\begin{document}

\title{Dynamical phase diagram of synchronization in one dimension: universal behavior from Edwards-Wilkinson to random deposition through Kardar-Parisi-Zhang}

\author{Ricardo Guti\'errez}
\author{Rodolfo Cuerno}
\affiliation{Universidad Carlos III de Madrid, Departamento de Matem\'aticas, Grupo Interdisciplinar de Sistemas Complejos (GISC), Avenida de la Universidad, 30 (edificio Sabatini), 28911 Leganés (Madrid), Spain}

\begin{abstract}
Synchronization in one dimension displays generic scale invariance with universal properties previously observed in surface kinetic roughening and the wider context of the Kardar-Parisi-Zhang (KPZ) universality class. This has been established for phase oscillators and also for some limit-cycle oscillators, both in the presence of columnar (quenched) disorder and of time-dependent noise, by extensive numerical simulations, and has been analytically motivated by continuum approximations in the strong oscillator coupling limit. The robustness and the precise boundaries in parameter space for such critical behavior remain unclear, however, which may preclude further developments, including the extension of these results to higher dimensions and the experimental observation of nonequilibrium criticality in synchronizing (e.g.~electronic or chemical) oscillators. We here present complete numerical phase diagrams of one-dimensional synchronization, including saturation times and values, but, most importantly, also dynamical features giving insight into the gradual emergence of synchronous dynamics, based on systems of phase oscillators with either type of randomness. In the absence of synchronization, the dynamics evolves as expected for random deposition (for time-dependent noise) or linear growth (for columnar disorder), while a crossover from Edwards-Wilkinson to Kardar-Parisi-Zhang behavior (with the corresponding type of randomness) is observed as the randomness strength, or the nonoddity of the coupling among oscillators, is increased in the synchronous region ---their combined effect being partially captured by the  so-called KPZ coupling. The distortion of scaling due to phase slips near the desynchronization boundary, a feature that is likely to play a role in experimental contexts, is also discussed.

\end{abstract}

\maketitle

\section{Introduction}

Synchronization, a ubiquitous form of collective dynamics arising in various scientific and technological contexts, has been a prominent research topic during the last four decades \cite{boccaletti_book, kuramoto_book,pikovsky}. Somewhat surprisingly, there is a strong relationship between this subject and the physics of surface kinetic roughening \cite{barabasi}, i.e.~the study of interfacial growth far from equilibrium, and, more widely, with Kardar-Parisi-Zhang (KPZ) universality \cite{takeuchi}. While the existence of a formal connection between synchronization models, on the one hand, and kinetic roughening and KPZ physics, on the other, has long been known at the level of the basic equations (see e.g.~\cite{kuramoto1984,pikovsky}), it is only in the last few years that a more quantitative relationship has been established, focused on critical exponents and general phenomenology \cite{lauter, moroney}.

Even more recently, in a series of works by the authors, the quantitative closeness between the two subjects has been elucidated to be considerably stronger than previously anticipated. One-dimensional (1D) systems of oscillators have thus been shown to display an array of universal features associated with universality classes previously studied in nonequilibrium critical dynamics \cite{PRR1,PRR2,PRE,PhysicaD}. The systems considered in those works include both phase-oscillators \cite{PRR1, PRE, PhysicaD} and limit-cycle oscillators \cite{PRR2, PRE}, under the presence of columnar disorder \cite{PRR1,PRR2} or time-dependent noise \cite{PRE}, and also a combination of both forms of randomness \cite{PhysicaD}. Extensive numerical simulations, backed by analytical arguments based on continuum approximations as well as phase-reduction calculations, show that the nonequilibrium criticality observed is quite robust and does not require setting parameters to specific critical values, representing instead an instance of generic scale invariance (GSI) whereby universal behavior emerges spontaneously and is robust throughout sizeable regions of parameter space  \cite{grinstein91,grinstein95,belitz05,tauber14}. For the dynamical systems and parameter values that have been inspected, the universal properties have been found to be those of the KPZ equation \cite{barabasi, takeuchi} with time-dependent \cite{PRE} or columnar disorder \cite{PRR1,PRR2}, depending on the type of randomness present in systems. The exception to the rule is that of coupling functions with odd symmetry \cite{PhysicaD} (a prime example being the celebrated Kuramoto model of phase oscillators \cite{acebron}), which results in an up-down symmetry of the phase-interface dynamics in the continuum, leading to large scale behavior in the universality class of the Edwards-Wilkinson (EW) equation \cite{barabasi} with the corresponding type of randomness. This has required studying different types of dynamic scaling in the phase correlations across time, which is Family-Vicsek \cite{vicsek} in the case of time-dependent noise \cite{PRE} and faceted anomalous \cite{ramasco} for columnar disorder \cite{PRR1}. And also investigating distributions of fluctuations around the average growth, which are of the Tracy-Widom (TW) type (see, e.g., \cite{takeuchi}) except for cases where the odd symmetry of coupling function leads to a Gaussian distribution \cite{PhysicaD}. When the coupling among oscillators is not sufficiently strong to overcome the effect of time-dependent noise, however, the dynamics follows the scaling of random deposition \cite{PRE}. For columnar disorder, on the other hand, nonsynchronous evolutions have not been studied systematically (as far as we know), yet linear growth may expected to be reached far away from the desynchronization boundary (as oscillators evolve at different mean effective frequencies) \cite{PRR1}.

While these universal features have been observed for multiple parameter choices, it is not fully clear how robust and general those types of critical behavior are. For example, the nonoddity of the coupling function (previously known to be relevant for the onset of synchronization, see Refs.~\cite{ostborn} and \cite{strogatz}) has consistenly been very high in the comparatively large systems studied in our previous work. That has been shown to enhance KPZ behavior, while smaller (nonzero) values of the nonoddity have been observed to lead to EW behavior, and something similar happens for fixed nonoddity when the system size is reduced \cite{PRR1,PRE}. Seemingly, this reflects the well-known EW-to-KPZ crossover behavior that has been studied in kinetic roughening \cite{forrest,Prolhac11}. Yet several questions arise in this regard. Is the nonequilibrium critical behavior uniform throughout the region in parameter space where synchronization is achieved for long times, or only in some portions of it? %How does it depend on the system size? 
What is the system behavior close to the desynchronization boundary, where continuum-approximation arguments presumably break down (as they require oscillator phases to satisfy a small-slope approximation \cite{PRR1})? 

In this paper, we place such recent results concerning the nonequilibrium criticality of 1D synchronization on a more secure footing by answering these questions through an extensive numerical study aimed at giving complete phase diagrams for phase oscillators in the presence of columnar disorder or time-dependent noise. In such diagrams, we display the following observables as functions of the randomness strength relative to the coupling strength and the nonoddity of the coupling function: 1) the saturation time (i.e.~the average time needed to achieve synchronization for a given parameter choice), 2) an order parameter quantifying the degree of synchronization achieved for long times, 3) the growth exponent of the phase interface in the emergence of synchronization, and 4) the skewness of the distribution of fluctuations around the average growth.  Observables 1) and 2) allow us to probe the (static, steady state) behavior of the system in terms of its propensity to synchronize. Most importantly from a dynamical perspective, observable 3) allows us to verify in which cases the growth corresponds to the exponent of the KPZ, EW or other universality classes, and observable 4) allows us to verify when the distribution is Gaussian or non-Gaussian, and in the latter case whether it appears to be TW or something else.

The structure of the paper is as follows. In the next section we discuss the model and observables under study. Then we present phase diagrams given in terms of those four observables, first for time-dependent noise, then for columnar disorder. Such diagrams allow us to partition parameter space into regions of qualitatively different critical behavior, whose shape and size appear to be partially captured by the so-called KPZ coupling \cite{barabasi,halpinhealy,takeuchi,Canet25}. After presenting the main conclusions derived from our numerical explorations, with special emphasis on their consequences for the experimental observability of critical behavior in one dimensional synchronization, we finally include four appendices that clarify some relatively technical aspects of our work. These include the estimation of saturation times, how various critical properties are affected in case the number of oscillators is relatively small,
%the role played by the system size on the critical behavior of synchronization, 
the distortion caused by phase slips near the desynchronization boundary, and the emergence of highly-skewed distributions around the same region under columnar disorder.

\section{Model, Observables, and Phase diagrams}

\subsection{Oscillator lattice model}

Consider a one-dimensional system of $L$ oscillators, where the state of oscillator $i\in \{1,2,\ldots, L\}$ at time $t$ is given by a phase variable $\phi_i(t) \in \mathbb{R}$. Such phase oscillators are idealized dissipative dynamical systems with an attracting limit cycle \cite{kuramoto_book}. Each of them interacts diffusively with its neighbors through a coupling function $\Gamma$ ---assumed to be smooth and $2\pi$-periodic in the relevant phase difference--- and is subject to some form of additive randomness $\eta_i$, resulting in the evolution equation
\begin{equation}
\frac{d \phi_i(t)}{d t} = \eta_i + \sum_{j\in \text{n.n.}} \Gamma[\phi_j(t) -\phi_i(t)].
\label{eqdisc}
\end{equation}
In the sum, $\text{n.n.}$ stands for $\{i-1,i+1\}$, using periodic boundary conditions (PBCs), i.e.~$\phi_0(t) \equiv \phi_{L}(t)$ and $\phi_{L+1}(t) \equiv \phi_1(t)$. The randomness term $\eta_i$ can take two forms: (i) time-dependent noise, $\eta_i \equiv \xi_i(t)$, which is white and Gaussian, and results in a stochastic (Langevin) evolution, or (ii) columnar disorder (see Ref.~\cite{PRR1} on motivation for the name), $\eta_i \equiv \omega_i$, i.e.~quenched noise given by a random assignment of natural frequencies $\omega_i$ (as in the Kuramoto model \cite{acebron}), resulting in a deterministic evolution. The combined effect of both types of randomness, not considered here, was explored in Ref.~\cite{PhysicaD}.

In either case, synchronization emerges when the coupling is strong enough to overcome the effect of the randomness, with all oscillators eventually attaining the same effective frequency, defined as
\begin{equation}
\omega^\text{eff}_i \equiv \lim_{T\to\infty}\frac{\phi_i(\tau+T)-\phi_i(\tau)}{T},
\label{omegaeff}
\end{equation}
after a time interval $[0,\tau]$ sufficiently long to contain the transient behavior (assuming such a limit exists). This kind of synchronization (less stringent than those definitions that require the asymptotic identity of the phases) is referred to as frequency locking or frequency entrainment in the literature.

\subsection{Coarse-grained description and KPZ equation}

A coarse-grained version of Eq.~(\ref{eqdisc}) is obtained by a continuum approximation previously developed in Refs.~\cite{PRR1,PRE}, which was originally adapted from Ref.~\cite{kuramoto_book}. Thus, the oscillators positions are continuous, $x\in \mathbb{R}$, so the phase of oscillator $i$, $\phi_i$, is denoted $\phi(x)$, and the neighboring oscillators are placed at positions  $x \pm a$, where $a$ is the lattice constant. Taylor expansions yield 
%$\phi(x \pm a,t) -\phi(x,t) =\pm a \partial_x \phi(x,t) + \frac{1}{2}a^2 \partial_x^2 \phi(x,t) + \mathcal{O}(a^3)$, and $\Gamma[\phi(x \pm a,t) -\phi(x,t)] = \Gamma(0) \pm a \Gamma^{(1)}(0) \partial_x \phi(x,t) + \frac{1}{2}a^2 \Gamma^{(1)}(0) \partial_x^2 \phi(x,t) + \frac{1}{2}a^2\Gamma^{(2)}(0) \left[\partial_x \phi(x,t)\right]^2 + \mathcal{O}(a^3)$, where $\Gamma^{(n)}(0)$ is the $n$-th derivative of the coupling function. Hence, Eq.~(\ref{eqdisc}) becomes
\begin{align}
\partial_t \phi(x,t) &= \eta(x,t) + 2 \Gamma(0) + a^2 \Gamma^{(1)}(0)\, \partial_x^2 \phi(x,t)\nonumber\\
&+ a^2 \Gamma^{(2)}(0) \left[\partial_x \phi(x,t)\right]^2 + \mathcal{O}(a^4).
\label{eqcont}
\end{align}
where $\Gamma^{(n)}(0)$ is the $n$-th derivative of the coupling function and $\eta(x,t)$ is a randomness (time-dependent noise or columnar disorder) term, including possible parameter renormalization. A constant term has been 
%proportional to $\Gamma(0)$ is inconsequential, and can be 
removed by considering oscillators in a co-moving frame, as typically done \cite{acebron}. Third-order terms are absent because (as for first and also higher odd-order terms) they vanish in the expansion due to the $x \to -x$ symmetry of the full coupling function $\Gamma$ (including both nearest neighbors).

Assuming small $a$, as compared to the scales over which $\phi(x)$ fluctuates, and neglecting $o(a^2)$ terms in Eq.~(\ref{eqcont}),
\begin{equation}
\partial_t \phi(x,t) =   \eta(x,t) + \nu \partial_x^2 \phi(x,t) + \frac{\lambda}{2}\left[\partial_x \phi(x,t)\right]^2.
\label{coarse}
\end{equation}
This is the KPZ equation, with either time-dependent \cite{kardar, barabasi} or columnar noise \cite{halpinhealy, szendro}, depending on the nature of the randomness $\eta(x,t)$. The parameters $\nu \equiv  a^2 \Gamma^{(1)}(0)$ and $\lambda/2 \equiv  a^2 \Gamma^{(2)}(0)$ follow the standard notation in the surface growth literature, where they quantify the surface tension and interface growth at a constant rate along the local surface normal direction, respectively \cite{barabasi,krug97}. In absence of the latter mechanism, $\lambda = 0$, the resulting linear evolution is known as the EW equation. %To make aspects of the discussion more intuitively clear, 
Henceforth, we will often describe the phase field $\phi(x,t)$ as a height profile over an oscillator substrate (each oscillator being located at a different $x$), as done in surface growth. A conspicuous difference between the two scenarios is the periodicity of the coupling function $\Gamma$, largely nonexistent for interfacial models but relevant close to the boundary of the synchronization region as discussed below.

The approximate description given by Eq.~(\ref{coarse}) has proven relevant to the large-scale behavior of the synchronization process as observed in numerical simulations based on both phase and limit-cycle oscillators \cite{PRR1,PRR2,PRE, PhysicaD}. Indeed, synchronization in such systems has been shown to possess robust universal features associated with the KPZ equation or the linear EW equation, both in the presence of columnar disorder \cite{PRR1,PRR2} and thermal noise \cite{PRE}. Note that the KPZ equation with columnar disorder, and its linear version for $\lambda = 0$, namely, the EW equation with columnar disorder (also known as the Larkin model \cite{purrello}), while less studied than the time-dependent-noise KPZ and EW equations \cite{barabasi,krug97,takeuchi}, have also been the focus of considerable efforts, and are characterized by distinct features very different from those of the  time-dependent-noise counterparts \cite{halpinhealy, szendro, purrello}.

An important property in the synchronization context is the symmetry of the coupling function under phase inversion, $\Delta \phi \to -\Delta \phi$ (where $\Delta \phi$ is the phase difference between two neighboring oscillators), particularly whether the function is odd, $\Gamma(\Delta \phi) + \Gamma(-\Delta \phi) = 0$, or not \cite{strogatz, ostborn}. This symmetry has been revealed crucial for several large-scale dynamical features of synchronization, being related to the occurrence of the nonlinearity in the continuum approximation, Eq.~(\ref{coarse}), i.e.\! whether $\lambda \neq 0$ \cite{PRR1,PRE,PhysicaD}. As $\lambda\propto\Gamma^{(2)}(0)$, indeed, if $\Gamma(\Delta \phi)$ is odd, so is its second derivative, which must vanish at the origin. As already clear in the oscillator model, Eq.\ \eqref{eqdisc}, only in this case does the system have up-down symmetry, i.e., invariance under phase reversal $\phi_i\to -\phi_i$, in a statistical sense, provided that the randomness distributions are (evenly) symmetric around their means \cite{PRR1}; in turn, Eq.\ \eqref{coarse} likewise is $\phi(x)\to -\phi(x)$ invariant thanks to $\lambda=0$.

\subsection{Coupling function and randomness}

Excluding higher harmonics in a Fourier expansion of the coupling function $\Gamma(\Delta \phi)$ yields the Kuramoto-Sakaguchi (KS) \cite{sakaguchi,sakaguchi86} coupling form
\begin{equation}
\Gamma(\Delta \phi) = K \sin(\Delta \phi + \delta),
\label{gamma}
\end{equation}
where $K$ is the coupling strength and $\delta \in (-\pi/2,\pi/2)$. The correspondence with the standard Fourier notation $\Gamma(\Delta \phi) = a_1\cos \Delta \phi +b_1 \sin \Delta \phi$ is given by $K = \sqrt{a_1^2 + b_1^2}$ and $\tan \delta = a_1/b_1$. The coupling is attracting, i.e.~$\Gamma^{(1)}(0) = K \cos \delta >0$, which amounts to positive surface tension $\nu = a^2 K \cos \delta$ in Eq.\ (\ref{coarse}). Moreover, $\lambda/2 = - a^2 K \sin \delta$, with the sign indicating only whether the KPZ nonlinearity drives growth in the local normal direction pointing upward (for $\delta < 0$) or downward (for $\delta >0$) \cite{barabasi, PRR1}. Thus $\tan \delta$, which we will refer to as the nonoddity of the coupling, %giving the relative strength of the $\cos \Delta \phi$ term to the $\sin \Delta \phi$ term, 
yields the relative strength (in absolute value) of the KPZ nonlinearity with respect to the surface tension. In fact, the odd symmetry $\Gamma(\Delta \phi) + \Gamma(-\Delta \phi) = 0$ mentioned above only holds for $\delta = 0$ ($\tan \delta = 0$), which correspond to $\Gamma(\Delta \phi) = K \sin \Delta \phi$ (Kuramoto coupling), yielding $\lambda = 0$ in Eq.\ (\ref{coarse}). The large-scale dynamics of the synchronization process in that case has indeed been shown to be in the universality class of the EW equation with columnar disorder \cite{PRR1} or time-dependent noise \cite{PRE}, as the case may be, in accordance with the absence of the KPZ term.

We next describe the randomness in the model, generically denoted as $\eta_i$ in Eq.~(\ref{eqdisc}). In the case of columnar disorder, $\eta_i=\omega_i$, the natural frequencies are taken to be independent and identically distributed according to a Gaussian with zero mean and standard deviation $\sigma_\text{col}$, i.e., $\langle \omega_i \rangle = 0$  and $\langle \omega_i \omega_j \rangle = \sigma_\text{col}^2\, \delta_{i j}$, where $\delta_{i j}$ is the Kronecker delta. As for time-dependent noise, $\eta_i = \xi_i(t)$, they are independent and Gaussian-distributed as well, with zero mean and standard deviation $\sigma_\text{tdep}$, and delta-correlated in time, $\langle \xi_i(t) \rangle = 0$, $\langle \xi_i (t)\, \xi_j (t')\rangle = \sigma_\text{tdep}^2\, \delta_{ij}\, \delta(t-t')$, where $\delta(\cdot)$ is the Dirac delta.

\subsection{Roughness and skewness}
\label{subsec:rough}

Here we discuss some observables, previously studied in the context of surface kinetic roughening \cite{barabasi,krug97,halpinhealy,takeuchi}, from which the quantities displayed in our phase diagrams for synchronziation will be derived. These are applied on phase profiles $\{\phi_i(t)\}_{i=1}^L$ resulting from simulations of the  spatially-discrete system in Eq.~(\ref{eqdisc}), with the coupling given in Eq.~(\ref{gamma}) and randomness as discussed at the end of the previous section. 
%In interfacial processes, such observables are applied on a height field $h(x,t)$ giving the local height of an interface growing above a one-dimensional substrate, and their adaptation to the study of phase fields $\phi(x,t)$ (or, rather, their discretization $\{\phi_i(t)\}_{i=1}^L$) has revealed novel aspects of the emergence of synchronization \cite{PRR1,PRR2,PRE,PhysicaD}. 

The spread of the local phases around the mean value is captured by the global width or roughness \cite{barabasi,halpinhealy,krug97}
\begin{equation}
\mathcal{W}(t) \equiv \langle \overline{[\phi_i(t)-\overline{\phi (t)}]^2} \rangle^{1/2},
\label{w}
\end{equation}
where the overbar denotes space average in a system of substrate size $L$ as $\overline{\phi(t)} = L^{-1} \sum_{i=1}^L \phi_i(t)$, and angular brackets denote averaging over different randomness realizations. As differences between oscillator phases that do not evolve at the same effective frequency $\omega^{\text{eff}}_i$ (\ref{omegaeff}) must grow steadily for long times, the saturation of $\mathcal{W}(t)$ to a time-independent value as $t\to\infty$ indicates synchronization in the sense of frequency locking mentioned above.

Critical dynamics in surface kinetic roughening implies that surface height values are statistically correlated for distances smaller than a correlation length $\xi(t)$ which increases with time as a power law, $\xi(t) \sim t^{1/z}$, where $z$ is the dynamic exponent \cite{barabasi}. The same correlation growth has been shown to be at play in the synchronization process between the phases of the oscillators distributed across space \cite{PRR1,PRR2, PRE}. Such an increase takes place until $\xi(t)$ reaches a value comparable to $L$, which happens at the saturation time $t_\text{sat} \sim L^z$ and results into the width saturating at a steady-state, size-dependent value $\mathcal{W}(t\gg L^z) \sim L^\alpha$. Here, $\alpha$ is the roughness exponent, which is related with the fractal dimension of the interface profile $\phi(x)$ \cite{barabasi,Mozo22}. In a wide variety of physical contexts and conditions (including classical models of equilibrium critical dynamics \cite{tauber14,Vaquero25}), the global roughness satisfies the Family-Vicsek (FV) dynamic scaling Ansatz
\cite{barabasi,halpinhealy,krug97,vicsek} $\mathcal{W}(t) = t^{\beta} f(L/\xi(t))$,
where the scaling function $f(y) \sim y^\alpha$ for $y\ll 1$, while $f(y) \sim {\rm cnst.}$ for $y\gg 1$. The ratio $\beta = \alpha/z$ is known as the growth exponent, and characterizes the short-time behavior of the roughness, $\mathcal{W}(t\ll L^z)\sim t^{\beta}$. The FV Ansatz is verified by important universality classes of kinetic roughening, like those of the KPZ and EW equations with time-dependent noise, being thus characterized by the set of $(\alpha,z)$ exponent values and their dependence on the substrate dimension $d$ \cite{barabasi,halpinhealy,krug97}. The values of the growth exponent in the presence of time-dependent noise, which will be relevant later, are $\beta_{\text{EW}} = 1/4$ and $\beta_{\text{KPZ}}  = 1/3$ \cite{barabasi}, see Table \ref{tab:exps}. Lattices of (both phase and limit-cycle) noisy oscillators have been recently shown to be in such universality classes (at least for some parameter choices), hence to display FV scaling \cite{PRE}. Additionally, for nonsynchronous dynamics in the presence of time-dependent noise the relevant universality class has been shown to be that of random deposition \cite{PRE}, with $\beta_{\text{RD}} = 1/2$ \cite{barabasi}.

The oscillator lattices with columnar disorder studied in Refs.~\cite{PRR1,PRR2} actually obey a related but different (i.e.\! non-FV) dynamic scaling ansatz termed anomalous scaling \cite{schroeder93,dassarma,krug97,lopez,ramasco,cuerno04}. While for standard FV systems height fluctuations at local distances $\ell \ll L$ scale with the same roughness exponent as global fluctuations do at distances comparable with the system size $L$, in systems displaying anomalous scaling local and global fluctuations scale with different roughness exponents, i.e. $w(\ell,t\gg \ell^z) \sim \ell^{\alpha_\text{loc}}$ with $\alpha_\text{loc} \neq \alpha$. The anomalous scaling that occurs in the synchronization process is most conveniently identified by means of the surface structure factor characterizing two-point correlations in Fourier space, as a new independent exponent $\alpha_s$ appears in the dominant contribution \cite{ramasco}. If $\alpha = \alpha_s >1$, as for the EW equation with columnar disorder \cite{purrello}, and observed in the synchronization of oscillators with a random assignment of intrinsinc frequencies when the coupling is odd \cite{PRR1,PhysicaD}, the anomalous scaling is termed super-rough \cite{lopez}, due to the large interface fluctuations that occur. The dynamic scaling ansatz satisfied by the structure factor is FV in this case, but $\alpha_\text{loc} = 1 \neq \alpha$. Otherwise, if $\alpha \neq \alpha_s$ with both exponents being larger than 1, another type of scaling termed faceted anomalous scaling takes place \cite{ramasco}, and again $\alpha_\text{loc} = 1$, as in the KPZ equation with columnar disorder \cite{szendro}, and seen for phase and limit-cycle oscillators in Refs.~\cite{PRR1,PRR2}. The growth exponents of the EW and KPZ equations with columnar disorder are $\beta_\text{cEW} = 3/4$ \cite{purrello} and $\beta_\text{cKPZ} = 1.07/1.37 \approx 0.78$, with the latter value, chosen after Ref.~\cite{PRR2}, being only approximate, as it is affected by non-universal corrections related to the disorder distribution \cite{krughh}. Such non-universal corrections must be reflected in our phase diagrams. For nonsynchronous dynamics under columnar disorder, linear growth (LG) of the roughness $\mathcal{W}(t)$ is expected, with $\beta_\text{LG} = 1$, due to different effective frequencies in the system \cite{PRR1}. 

For the reader's convenience, Table \ref{tab:exps} collects the relevant values of the critical exponents just described. See also Appendix \ref{appsat} for some analytical derivations, useful to estimate saturation times. Specifically, in the phase diagrams below $\beta$ will be directly probed. It is actually the only available exponent in cases like RD and LG behavior, where the space correlation length is not well defined. 

\begin{table}[]
    \centering
    %\footnotesize %\small
    \begin{tabular}{|c||c|c|c|c|}
    \hline
    Universality class & $\alpha$ & $\alpha_s$ & $\beta$ & $z$ \\
    \hline \hline
       RD (time dep.\ noise) \cite{barabasi} & $-$ & $-$ & 1/2 & $-$ \\
       \hline
       LG (columnar disorder) & $-$ & $-$ & 1 & $-$ \\
       \hline
       EW (time dep.\ noise) \cite{barabasi} & 1/2 & 1/2 & 1/4 & 2 \\
       \hline
       cEW (columnar disorder) \cite{purrello,PRR1} & 3/2 & 3/2 & 3/4 & 2 \\
       \hline
       KPZ (time dep.\ noise) \cite{barabasi} & 1/2 & 1/2 & 1/3 & 3/2 \\
       \hline
       cKPZ (columnar disorder) \cite{szendro,PRR1} & 1.07 & 1.40 & 0.79 & 1.36 \\
    \hline
    \end{tabular}
    \caption{Kinetic roughening exponent values relevant to this work. Values are exact if rational and numerically estimated otherwise. ``$-$'' denotes undefined. See also Appendix \ref{appsat}.}
    \label{tab:exps}
\end{table}

Beyond averaged quantities like $\mathcal{W}(t)$ (\ref{w}), an observable of interest that has received much attention during the last decade is the probability distribution function (PDF) of the fluctuations of the local field around its mean, in our case $\delta \phi_i(t) = \phi_i(t) - \overline{\phi(t)}$. After normalizing by the systematic increase given by $\mathcal{W}(t) \sim t^\beta$, such PDF reaches a universal, time-independent form \cite{kriecherbauer10,halpinhealy,takeuchi}.
Important examples in the kinetic roughening literature are, e.g., the Gaussian distribution for the linear EW equation \cite{barabasi,krug97} and a TW PDF (whose precise form depends, e.g., on boundary conditions) for the KPZ equation \cite{kriecherbauer10,halpinhealy,takeuchi}, both of them with time-dependent noise. Except for Kuramoto coupling, which is associated with Gaussian fluctuations, all results contained in Refs.~\cite{PRR1,PRR2,PRE} for generic coupling functions which do not possess odd symmetry (including those with columnar disorder) show TW fluctuations associated with the Gaussian Orthogonal Ensemble (GOE) of random matrix theory, as expected for the 1D KPZ equation with time-dependent noise and PBC \cite{kriecherbauer10,takeuchi}. For limit-cycle oscillators, the absence of such odd symmetry from the coupling between phases in their phase-reduced approximations \cite{pietras} has been confirmed in some cases \cite{PRR2}, and is expected to hold quite generally, even more so if higher-order terms are considered. Characterizing these fluctuations is quite demanding computationally, as large system sizes on the order of several thousands of oscillators are required to achieve a good characterization of the distribution.

To quantify how (non-)Gaussian is the fluctuation PDF in our numerical results, we use the skewness,
\begin{equation}
\mathcal{S}(t) = \frac{\langle \overline{[\phi_i(t)-\overline{\phi (t)}]^3} \rangle}{\langle \overline{[\phi_i(t)-\overline{\phi (t)}]^2} \rangle^{3/2}},
\label{skew}
\end{equation}
which is zero for a Gaussian PDF and approximately $\mathcal{S}_\text{TW} = 0.29346452408$ for GOE-TW \cite{Bornemann10}. The zero value is expected to hold for synchronizing oscillators with either type of randomness and odd coupling \cite{PhysicaD}, while the GOE-TW value is expected for couplings where the nonoddity of the function $\Gamma(\Delta \phi)$ leads to a prominent KPZ nonlinearity, at least for sufficiently large system sizes \cite{PRR1,PRR2,PRE,PhysicaD}. Note also that, for the KPZ equation, the sign of the skewness of the fluctuation distribution is that of the nonlinear parameter $\lambda$ in Eq.\ \eqref{coarse} \cite{takeuchi}. For nonsynchronous dynamics, fluctuation PDFs have not been studied as far as we are aware, yet virtually uncoupled oscillators are expected \cite{Marcos25} to undergo Gaussian fluctuations under time-dependent noise of sufficient strength.

\subsection{Simulations and saturation time estimation}
\label{simsat}

Our phase diagrams for 1D synchronization are based on simulations of rings of KS oscillators, Eqs.~(\ref{eqdisc}, \ref{gamma}), starting from a flat initial condition, $\phi_i(0)=0$ for all $i$. The evolution is implemented numerically with a time step $\delta t = 0.01$, through an Euler-Mayurama scheme \cite{Toral} in the stochastic case of time-dependent noise, and through a standard fourth-order Runge-Kutta algorithm in the deterministic case of columnar disorder. The main results to be displayed in the next section are for rings of $L = 1000$, and are based on hundreds of independent realizations. %(i.e.~noise histories for time-dependent noise; random assignments of intrinsic frequencies for columnar disorder). 
Given that, e.g., experimental constraints may substantially constrain the maximum available number of interacting oscillators, results for smaller systems are included in Appendix \ref{appsmall} to illustrate how universal behavior is affected when restricting scaling regimes. 

As phase diagrams are given in terms of observables derived below from the roughness and the skewness, in our simulations we store values for $\mathcal{W}(t)$ (\ref{w}) and $\mathcal{S}(t)$ (\ref{skew}) at logarithmically equispaced time points $t= 1, 1.5, 1.5^2 = 2.25, 1.5^3 = 3.375$ and so on, up to a given maximum time $t^M$ corresponding to some power of the multiplicative factor $1.5$. The length $t^M$ of the time window is chosen large enough so that, if saturation has not happened by then, it is unlikely to take place. A consistency check on this choice of $t^M$ consists in inspecting the phase diagram observables obtained for the corresponding numerical data, as they lead to well-defined behavior for nonsynchronous dynamics that differs strongly from that observed for synchronous dynamics, see the next section for more details. If the system has not saturated by $t=t^M$, yet displays the hallmarks of synchronous dynamics, that would imply that $t^M$ is too short and must be increased. Several iterations of this procedure serve to find suitable values of $t^M$ (while not being so large as to be computationally wasteful), within our numerical precision.

We next focus on the numerical estimate, $t^*$, of the saturation time of the roughness, $t_\text{sat}$, which is basic for our phase diagrams. As the procedure starts from the last time point $t^M$, for convenience we denote points appearing earlier in a sequence as $t^M_n = t^M/1.5^n$, with $n \in \mathbb{N}$ (hence, $t^M_1 = t^M/1.5$ is one time point before $t^M$, $t^M_2 = t^M/1.5^2$ two points before $t^M$, etc.). To determine $t^*$, we start from $t=t^M$ (an upper bound for $t^*$ by construction) and check that the smallest difference (in absolute value) between the roughness at $t^M$ and at the previous three points ($t^M_1, t^M_2, t^M_3$) is less than 10\% of its value at the reference time $t^M$, namely,
$\min_{n=1,2,3}\{\,|\mathcal{W}(t^M)-\mathcal{W}(t^M_n)| \,\} < 0.1\, \mathcal{W}(t^M)$. %,|\mathcal{W}(t^M)-\mathcal{W}(t^M_2)|,|\mathcal{W}(t^M)-\mathcal{W}(t^M_3)|\} < 0.1\, \mathcal{W}(t^M)
If that is the case, we consider that we are still at saturation and move to one point earlier in the sequence, namely $t^M_1$, and compare the roughness at that new reference point with the values at $t^M_2, t^M_3$, and $t^M_4$. If the saturation condition is still satisfied, we take $t^M_2$ as our reference point and proceed analogously, continuing until the condition fails. Once this happens, the procedure stops and we take the saturation time estimate $t^*$ to be that reference time point where the condition failed, i.e., the first in the sequence (starting from $t^M$ and proceding backwards) for which all three relative roughness differences with respect to the three previous points are larger than $10\%$. If the roughness values for a given parameter choice violate the saturation criterion already for $t=t^M$, saturation never occurs, yet we take $t^* = t^M$ in practice for the sake of computing the observables to be described in the next subsection, which require assigning some value to $t^*$.

The reason for considering three roughness differences (and not just one) in the saturation criterion is just to increase its robustness against statistical fluctuations, whose existence, together with the sparseness of our (logarithmically-equispaced) time points, due to computational reasons, imply that the most we can realistically expect is that $t^*$ is of the same order of magnitude as the actual $t_\text{sat}$. Yet we find that our estimate $t^*$ does provide very reasonable results, at least in logarithmic scale. % (which is the natural setting for a discussion of critical properties).
The detailed results reported in Appendix \ref{appsat} illustrate this in several cases of interest, including comparisons between our $t^*$ and theoretical estimates of the saturation time $t_\text{sat}$ well established in the kinetic roughening literature.

Before moving on to discussing the phase diagrams proper, we introduce an additional piece of notation by analogy with that employed to denote points to the left of $t^M$. Namely, we denote time points appearing earlier than $t^*$ in the sequence as $t^*_n = t^*/1.5^n$. Especially important in the following will be $t^*_{5} \approx  0.13 \, t^*$ and $t^*_{11} = t^* \approx 0.012\, t^*$, roughly one and two orders of magnitude below the estimated saturation time $t^*$. That is so because  $[t^*_{11},t^*_5]$ will be the reference interval within the growth regime where we assume that $\mathcal{W}(t) \sim t^\beta$ holds in the phase diagrams aimed to characterize dynamical features of the growth process. When saturation is not reached, $t^* = t^M$, the interval $[t^*_{11},t^*_5]$ is expected to encompass long enough times for the system to be showing its representative asymptotic behavior for a nonsynchronous evolution.

\subsection{Phase diagrams}
\label{subsec:PD}

Our phase diagrams represent certain observables as functions of two control parameters:
\begin{itemize}
\item[(i)] $\tan \delta$: the nonoddity of the coupling function;
\item[(ii)] $D$: the relative strength of randomness vs coupling.
\end{itemize}
The physical meaning of (i) has been elucidated right after presenting the KS coupling in Eq.~(\ref{gamma}). As to (ii), it is given by the ratio of the noise strength to the coupling strength, $D = \sigma_\text{tdep}/K$ or $\sigma_\text{col}/K$, depending on the type of randomness. For a given coupling function $\Gamma$ (set by the choice of $\delta$ or, equivalently, by the nonoddity $\tan \delta$ in the case of KS coupling), this adimensional parameter is the only relevant one for the synchronization problem \cite{pikovsky}. In fact, generally speaking, it is the relative strength of terms on the right hand side of Eq.~(\ref{eqdisc}) that matters, as varying the absolute strengths amounts to a redefinition of the time unit. For simplicity, we fix the overall coupling strength $K=1$ and just take $D =\sigma_\text{tdep}$ or $\sigma_\text{col}$, depending on the randomness under consideration. 

We run independent simulations for values of the nonoddity $\tan \delta$ in the interval $[-10,10]$, and for values of $D$ in the interval $[0,0.8]$ (for time-dependent noise) or $[0, 0.4]$ (for columnar disorder). The ranges of $D$ are simply chosen to be large enough for the phase diagram to accommodate the parameter-space region showing synchronization, as well as a substantial part of the nonsynchronous region. That this is achieved for smaller $D$ in the case of columnar disorder is likely related to the persistence in time of that form of randomness, as opposed to the lack of persistence of our time-dependent noise.

The observables themselves, shown in the phase diagrams as functions of $\tan \delta$ and $D$, are:
\begin{itemize}
\item[(A)] The saturation time normalized by the maximum time, $t^*/t^M$. By definition, $t^*/t^M \leq 1$, equality indicating that saturation (as given by the numerical condition discussed above) does not occur. 
\item[(B)] The roughness at saturation normalized by the nonsynchronous power-law growth $\mathcal{W}(t^*)/(t^M)^{\beta_\text{div}}$, with $\beta_\text{div} = \beta_\text{RD} = 1/2$ for time-dependent noise and $\beta_\text{div} = \beta_\text{LG} = 1$ for columnar disorder, see Table \ref{tab:exps}. As $\mathcal{W}(t)$ is bounded for $t\geq t^*$ under synchronization, while it diverges as $t^{\beta_\text{div}}$ for nonsynchronous dynamics, this observable will be on the order of $1$ in the latter case and much smaller in the former.
\item[(C)] An estimate of the growth exponent given by
\begin{equation}
\beta^* = \frac{\log[\mathcal{W}(t^*_{5})/\mathcal{W}(t^*_{11})]}{\log[t^*_{5}/t^*_{11}]},
\label{betastar}
\end{equation}
where we are assuming $\mathcal{W}(t) \sim t^\beta$. This growth is uninterrupted under nonsynchronous conditions as discussed for (B), with exponent $\beta = \beta_\text{div}$. In the case of synchronization, $\beta$ may take any of the other values listed in Table \ref{tab:exps} (with possible modifications due to finite-size effects, non-universal dependencies, etc.) while growth eventually saturates. 
\item[(D)] An averaged skewness across the growth regime,
%An estimate of the skewness averaged throughout the growth regime,
\begin{equation}
\mathcal{S}^* = \frac{1}{7} \sum_{n=5}^{11} \mathcal{S}(t^*_n).
\label{sstar}
\end{equation}
\end{itemize}

Representing $t^*/t^M$ (A) and $\mathcal{W}(t^*)/(t^M)^{\beta_\text{div}}$ (B) as functions of $\tan \delta$ and $D$ gives a static (steady-state) phase diagram of synchronization. For either type of randomness, those two diagrams will be shown first  in order to answer the following question: {\it For which $\tan \delta$ and $D$ does the system synchronize at long enough times?}

On the other hand, $\beta^*$ (C) and $\mathcal{S}^*$ (D) as functions of the same control parameters yield two complementary dynamical phase diagrams concerning the growth regime, where synchronization is still emerging (in case of saturation) and $\mathcal{W}(t) \sim t^\beta$. These are our main objects of interest. The questions addressed are of a very different nature in those diagrams, namely: {\it What is the dynamical process that leads towards synchronization for different parameter choices? Does it correspond to any universal behavior previously observed in nonequilibrium critical dynamics? How prevalent is such GSI behavior, previously reported in Refs.~\cite{PRR1,PRR2,PRE,PhysicaD}, in parameter space?}

When saturation does not occur for certain choices of $(\tan \delta, D)$, we find $t^*/t^M = 1$ in the first phase diagram (A). For those parameter values, assuming they do reflect a true nonsynchronous evolution, we expect $\mathcal{W}(t^*)/(t^M)^{\beta_\text{div}}$ to be on the order of $1$ in (B), $\beta^* \approx 
 \beta_\text{RD} = 1/2$ (for time-dependent noise) or $\beta^* 
 \approx \beta_\text{LG} = 1$ (for columnar disorder) in (C), and $S^* \approx 0$ for time-dependent noise in (D). (We will see that $S^*$ for nonsynchronous dynamics in the presence of columnar disorder has a much more complex behavior.) Significantly different values in phase diagrams (B), (C), and (D), corresponding to behaviors consistently different from RD or LG, for those parameter-space points corresponding to $t^*/t^M = 1$ in (A) suggest that $t^M$ needs to be increased, as the dynamics is potentially synchronous, only it did not have enough time to saturate. This is how the consistency check mentioned above has been implemented, resulting in a $t^M$ for which such inconsistent values in the phase diagrams have been removed.

 One last theoretical aspect of our work must be discussed before we present our numerical results. In our phase diagrams we will see that both increasing the randomness strength $D$ (trivially) and increasing  the nonoddity $\tan \delta$ (perhaps less so, yet see Ref.~\cite{lauter}) eventually lead to desynchronization. But the synchronous region is far from being a rectangle in parameter space: there does seem to be some interrelation between the effects of the two parameters, in the sense the value of $\tan \delta$ that leads to desynchronization for fixed $D$ depends on the specific value of $D$, and the other way around. Moreover, there are regions of parameter space that display similar behaviors more generally (not only in connection with the desynchronization boundary), whereby a given $\tan \delta$ for a given $D$ appears to have equivalent effects to a smaller $\tan \delta$ for a larger $D$. In this regard, we have found that the effective KPZ coupling \cite{barabasi,halpinhealy,takeuchi,Canet25}, defined (up to a dimension-dependent constant factor) as
 \begin{equation}
 g = \frac{\lambda^2 \mathcal{D}}{\nu^3},
 \label{KPZcoupl}
 \end{equation}
 where $\mathcal{D}$ plays the role of $D^2/2$ here,%(for our microscopic noise strength $D$)
 \footnote{Equation \eqref{KPZcoupl} defines the effective KPZ coupling for the KPZ equation \eqref{coarse}, where $\langle \eta(x,t)\eta(x',t')\rangle = 2\mathcal{D}\delta(x-x')\delta(t-t')$.} seems to capture %at least some of 
 this dependence between the parameters $\tan \delta$ and $D$, for time-dependent noise but also (somewhat surprisingly) for columnar disorder. It turns out that $g$ governs the renormalization-group flow of the KPZ equation with time-dependent noise, thus controlling its emergent scaling behavior at large space and time scales, yielding EW for small $g$ values (weak coupling) and KPZ for large $g$ (strong coupling) \cite{barabasi,Canet25}. According to the discussion of the continuum-approximation parameters for the KS model following Eq.~(\ref{gamma}), we use as our (na\"ive) proxy to this renormalized effective parameter the following variable depending on microscopic (bare) parameters alone:
 \begin{equation}
  g^* = \frac{D^2 \tan^2{\delta}}{\cos \delta}.
 \label{gstar}    
 \end{equation}
We have removed factors which are purely numerical or that depend on the lattice spacing $a$, as we will only be interested in relative values of $g^*$ across parameter $(\tan \delta, D)$-space. Note that $\cos \delta$ in the denominator of Eq.~(\ref{gstar}) is calculated in terms of the nonoddity $\tan \delta$ as $\cos(\arctan({\tan \delta}))$ without ambiguity, as $\delta \in (-\pi/2,\pi/2)$; moreover, $g^*$ is invariant under $\delta \leftrightarrow -\delta$.

In the phase diagrams of the next section, the approximate KPZ coupling $g^*$ will be used to delineate regions in parameter space showing a similar scaling behavior. Note that the use of the same denomination of ``coupling'' for both $g$ (\ref{KPZcoupl}) and $K$ (\ref{gamma}), follows standard practice in two different physical backgrounds (renormalization group vs coupled dynamical systems), but the two constants should not be confused or taken to be equivalent. In fact, the microscopic approximation to the KPZ effective coupling $g^*$ is inversely proportional to the oscillator coupling $K$; we have set $K=1$ in Eq.~(\ref{gstar}), as that is the value of choice for the rest of this work.

\section{Results}

This section contains phase diagrams for 1D synchronization, in terms of the observables and control parameters described above. A key aspect of the oscillator model employed, with strong qualitative repercussions in those diagrams as it affects the emergence of synchronization and the form of GSI that it takes, is the nature of the randomness. Accordingly, we split the presentation and discussion of numerical results into two subsections.%, with the first one focusing on time-dependent noise and the second one on columnar disorder.

\subsection{Time-dependent noise}

\begin{figure}[t!]
\includegraphics[scale=0.42]{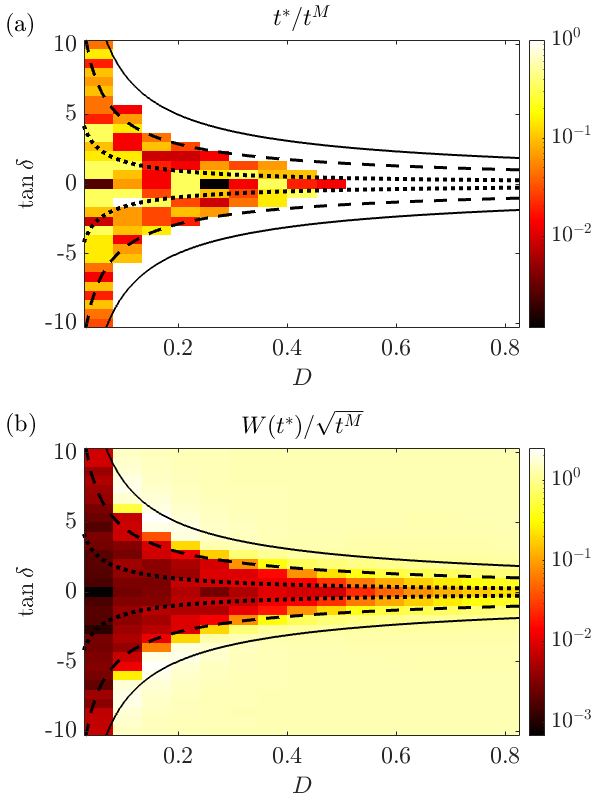}
\caption{{\sf \bf Static phase diagrams for rings of $L=1000$ KS oscillators in the presence of time-dependent noise.} (a) Normalized estimate of the saturation time $t^*/t^M$ as a function of the non-oddity $\tan \delta$ and the noise strength $D$. (b) Normalized roughness at saturation $\mathcal{W}(t^*)/\sqrt{t^M}$ as a function of the same two parameters. The solid/dashed/dotted black lines show $g^* = 5/1/0.05$ level curves, cf.\ Eq.\ \eqref{gstar}. See Sec.~\ref{subsec:PD} for additional definitions and expected behaviors. Results based on 500 realizations.}
\label{PDstat_td}
\end{figure}

In this subsection we show phase diagrams for rings of KS oscillators in the presence of time-dependent noise. We first focus on diagrams that shed light on the existence of synchronization for long times, with a wide range of values for the nonoddity $\tan \delta$ and the noise strength $D$ explored, so as to encompass synchronous as well as nonsynchronous dynamics. Specifically, Fig.~\ref{PDstat_td} shows the normalized saturation time $t^*/t^M$ [panel (a)] and the roughness at saturation $\mathcal{W}(t^*)/(t^M)^{\beta_\text{div}} = \mathcal{W}(t^*)/\sqrt{t^M}$ [panel (b); here $\beta_\text{div} = \beta_\text{RD} = 1/2$] as a function of those two control parameters. Our values for $t^*$ provide reasonable order-of-magnitude estimates of the saturation time, see Appendix \ref{appsat} for details, including comparison with theoretical values previously reported in the literature.
%For further discussion of the saturation time estimate $t^*$ and its comparison with theoretical values previously reported in the literature, see Appendix \ref{appsat}, which shows that $t^*$ provides reasonable order-of-magnitude estimates. 

The diagrams in Fig.~\ref{PDstat_td} mostly serve the purpose of first delineating regions in parameter $(\tan \delta, D)$-space where synchronization takes place with our particular choice of model and parameters. %This will be key in order to study the GSI that is associated to its emergence. 
Thus, for large enough $\tan \delta$ or $D$, the system displays nonsynchronous dynamics, specifically in the region where $t^* = t^M$, which also corresponds to a roughness $\mathcal{W}(t^*)/\sqrt{t^M} \approx 1$, large enough to be compatible with the absence of saturation. Further evidence based on dynamical considerations is provided below.

Synchronization is seen to be lost when the noise strength $D$ increases (as expected), but also when the nonoddity $\tan \delta$ does, which corresponds to a increased KPZ nonlinearity relative to the surface tension, from the point of view of Eq.\ \eqref{coarse}. The desynchronization boundary displays a nontrivial dependence on those control parameters, which the approximate KPZ coupling $g^*$ (\ref{gstar}) seems to capture, as shown by its level curves for $g^* = 5$ (continuous black lines), $1$ (dashed black lines) and $0.05$ (dotted black lines) in Fig.~\ref{PDstat_td}. For values of $g^*$ around $g^* = 5$ and larger synchronization is absent. But it is the regions with smaller $g^*$, yet not too small (with relatively large $\tan \delta$ approaching the desynchronization boundary from within), that will be of greatest interest in the dynamical phase diagrams, which is the subject we turn to next.

\begin{figure}[t!]
\includegraphics[scale=0.42]{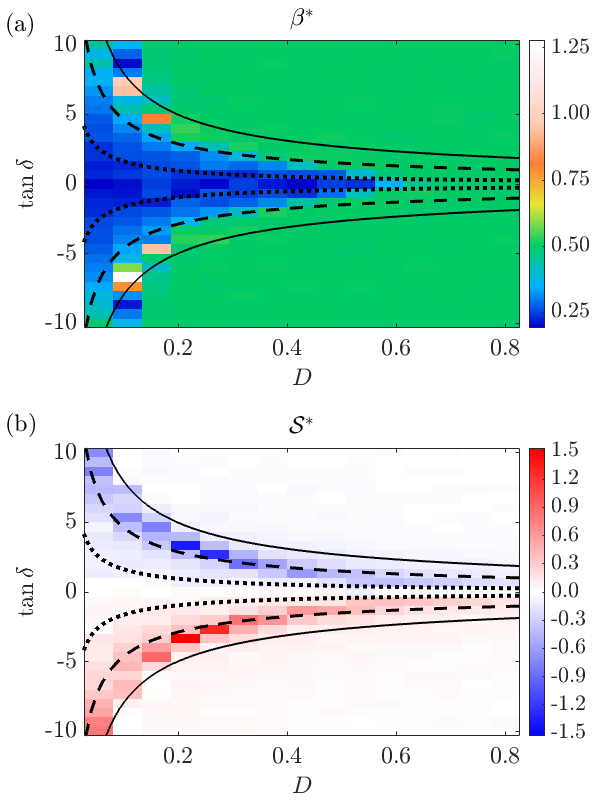}
\caption{{\sf \bf Dynamic phase diagrams for rings of $L=1000$ KS oscillators in the presence of time-dependent noise.} (a) Estimate of growth exponent $\beta^*$ as a function of the non-oddity $\tan \delta$ and the noise strength $D$. (b) Estimate of skewness of fluctuations $\mathcal{S}^*$ as a function of the same two parameters. The solid/dashed/dotted black lines show $g^* = 5/1/0.05$  level curves, cf.\ Eq.\ \eqref{gstar}. See Sec.~\ref{subsec:PD} for additional definitions and expected behaviors. Results based on 500 realizations.}
\label{PDdyn_td}
\end{figure}

In Fig.~\ref{PDdyn_td} we show the estimate of the growth exponent $\beta^*$ [panel (a)] and of the skewness $\mathcal{S}^*$ [panel (b)] as functions of $\tan \delta$ and $D$, including the same level curves for $g^*$ displayed in Fig.~\ref{PDstat_td}. In those regions where that previous figure shows nonsynchronous dynamics, we now find $\beta^* \approx \beta_\text{RD} = 1/2$ and $S^* \approx 0$, as expected. In the presence of synchronization the situation is more complex. For very small $g^*$ we find values compatible with EW universality (previously reported for $\delta = 0$ \cite{PRE}  and other odd couplings \cite{PhysicaD}), namely $\beta^* \approx \beta_\text{EW} = 1/4$ and $S^* \approx 0$. Yet for larger $g^*$ (see the region comprised between the two level curves for $g^* =0.05$ and $1$) the dynamics is still synchronous, but with a clearly non-EW behavior. The values of $\beta^*$ there are mostly compatible with $\beta_\text{KPZ} = 1/3$. Moreover, there is a clearly nonzero $\mathcal{S}^*$ which appears to be compatible with the TW value ($\mathcal{S}_\text{TW} \approx 0.29$), which also suggests KPZ behavior as previously reported in Refs.~\cite{lauter} and \cite{PRE}. For larger values of $g^*$, roughly in between the level curves for $g^* = 1$ and $5$, the behavior is less clear, and is frequently characterized by larger $S^*$ values (occasionally $\beta^*>1/2$), see also below.

While the results reported in Fig.~\ref{PDstat_td} for a given $D$ are not systematically affected by a change of sign in $\delta$, %(hence that in the nonoddity $\tan \delta$)
in Fig.~\ref{PDdyn_td} we observe that likewise $\beta^*$ does not change significantly, whereas $S^*$ undergoes a clear sign reversal, being negative for $\delta >0$ and positive for $\delta < 0$. This must be related to the fact that $\delta \to - \delta$ leads to a sign flip in the cosine term in the KS coupling (\ref{gamma}), and hence in the KPZ nonlinearity in a coarse-grained description (\ref{coarse}). Moreover, the average velocity of the phase profile also changes sign (not shown), from positive for $\delta >0$ to negative for $\delta <0$, as can be deduced from the fact that
\begin{align}
\overline{\frac{d \phi_i}{d t}} &= \overline{\xi_i }+  K[ \overline{\sin(\phi_{i+1}\!-\!\phi_i\!+\!\delta)\!+\!\sin(\phi_{i-1}\!-\!\phi_{i}\!+\!\delta)}]\nonumber \\
&\approx K[ \overline{\sin(\phi_{i+1}\!-\!\phi_i\!+\!\delta)}\!+\!\overline{\sin(\phi_{i}\!-\!\phi_{i+1}\!+\!\delta)}]\nonumber\\
&\approx 2 K \sin \delta\, \overline{\cos (\phi_{i+1}- \phi_i)},
\label{avfreq}
\end{align}
where we have applied PBC, and the approximate equality arises from assuming that the system is large enough so that the law of large numbers yields $\overline{\xi_i } \approx \langle \xi_i \rangle = 0$.

Phase diagrams in Figs.~\ref{PDstat_td} and \ref{PDdyn_td} correspond to rings of $L=1000$ oscillators, but other system sizes, including $L= 200$ and $500$, have also been inspected, with no substantial qualitative differences being observed (not shown). To illustrate this point, in Appendix \ref{appsmall} we show results for a smaller ring containing $L=100$ oscillators, which look quite similar to those reported here for $L=1000$, except for a larger EW-dominated region, and considerably increased absolute values in the skewness estimate $S^*$. This seems to be related to the observation \cite{PRE} that very large system sizes are required for unambiguously recovering the TW fluctuation PDF.

Regarding those points close to the boundary between synchronization and desynchronization in Fig.~\ref{PDdyn_td} where values of $\beta^*$ and $\mathcal{S}^*$ deviate considerably from $\beta_\text{KPZ}$ and $\mathcal{S}_\text{TW}$ (including some for which $\beta^*>1$), this seems to be related to the appearance of $2\pi$-phase slips (see, e.g., Ref.~\cite{pikovsky}), which occur near the desynchronization boundary, and also (trivially) over the nonsynchronous region. See Appendix \ref{appslips} for a detailed discussion of these deviations, including several numerical results that clarify aspects of the underlying dynamics.

In conclusion, the KPZ behavior for one-dimensional oscillators under time-dependent noise discussed in Refs.~\cite{PRE, PhysicaD} appears to be relatively delicate: for moderate $D$, if $\tan \delta$ is not large enough, there is a crossover to the EW behavior observed in the synchronous region for $\delta =0$, which is especially conspicuous for smaller system sizes, see Appendix \ref{appsmall}. As $\tan \delta$ or $D$ (or their combination as given in $g^*$) is increased, however, the system eventually abandons the EW region, and show signatures of KPZ universality. But, if those parameters keep being increased, soon enough the system is close to the desynchronization boundary, in a region where the appearance of such phase slips distorts the phase profile. This necessarily blurs the KPZ universal features expected to be observed at large space-time scales.

\subsection{Columnar disorder}

\begin{figure}[t!]
\includegraphics[scale=0.42]{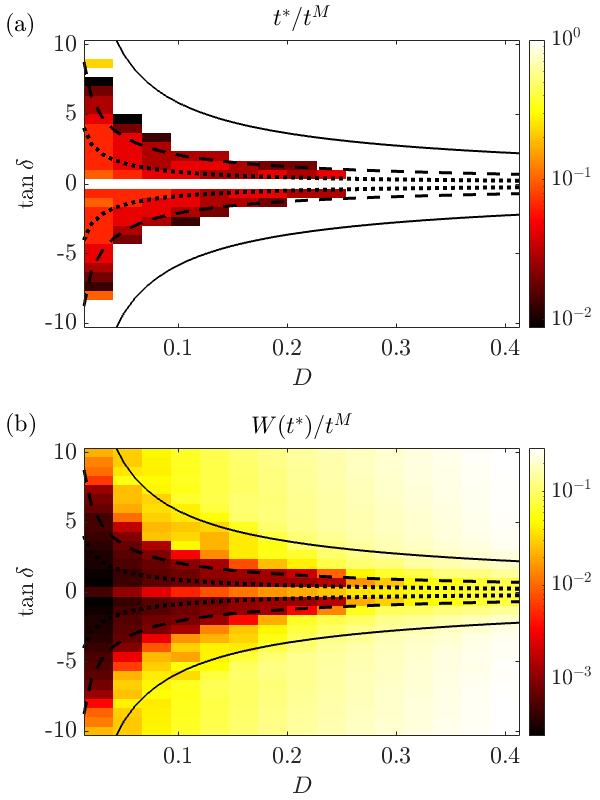}
\caption{{\sf \bf Static phase diagrams for rings of $L=1000$ KS oscillators in the presence of columnar disorder.} (a) Normalized estimate of the saturation time $t^*/t^M$ as a function of the non-oddity $\tan \delta$ and the noise strength $D$. (b) Normalized roughness at saturation $\mathcal{W}(t^*)/t^M$ as a function of the same two parameters. The solid/dashed/dotted black lines show $g^* = 2/0.1/0.01$ level curves, cf.\ Eqs.\ \eqref{KPZcoupl} and \eqref{gstar}. See Sec.~\ref{subsec:PD} for additional definitions and expected behaviors. Results based on 500 realizations.}
\label{PDstat_col}
\end{figure}

We next show phase diagrams for rings of KS oscillators in the presence of columnar disorder. The discussion in this section parallels that of the previous one for time-dependent noise, hence we will mostly focus on the new features that arise for this type of quenched randomness. Figure \ref{PDstat_col} shows the normalized saturation time $t^*/t^M$ [panel (a)] and the roughness at saturation $\mathcal{W}(t^*)/(t^M)^{\beta_\text{div}} = \mathcal{W}(t^*)/t^M$ [panel (b); here $\beta_\text{div} = \beta_\text{LG} = 1$] as functions of $\tan \delta$ and $D$. For further discussion of the saturation time estimate $t^*$ and its comparison with theoretical values, see Appendix \ref{appsat}, where $t^*$ is seen to provide reasonable order-of-magnitude estimates in the case of columnar disorder as well.

We observe that, for large enough $\tan \delta$ or $D$, the system displays nonsynchronous dynamics, specifically in the region where $t^* = t^M$, which also corresponds to a roughness $\mathcal{W}(t^*)/t^M$ only moderately smaller than $1$. The desynchronization boundary displays a nontrivial dependence on the control parameters, which again the approximate KPZ coupling $g^*$ (\ref{gstar}) seems to capture, as illustrated by the level curves shown for $g^* = 2$
(continuous black lines), $0.1$ (dashed black lines) and $0.01$
(dotted black lines). A distinct feature of the columnar-disorder case is that a very small disorder strength $D$ is enough to suppress synchronization in the particular case of $\delta = 0$. In fact, this result is linked to the proven absence of synchronization for odd coupling in the thermodynamic limit \cite{strogatz}, whose consequences for finite sizes were previously probed numerically in Ref.~\cite{PRR1}.

\begin{figure}[t!]
\includegraphics[scale=0.42]{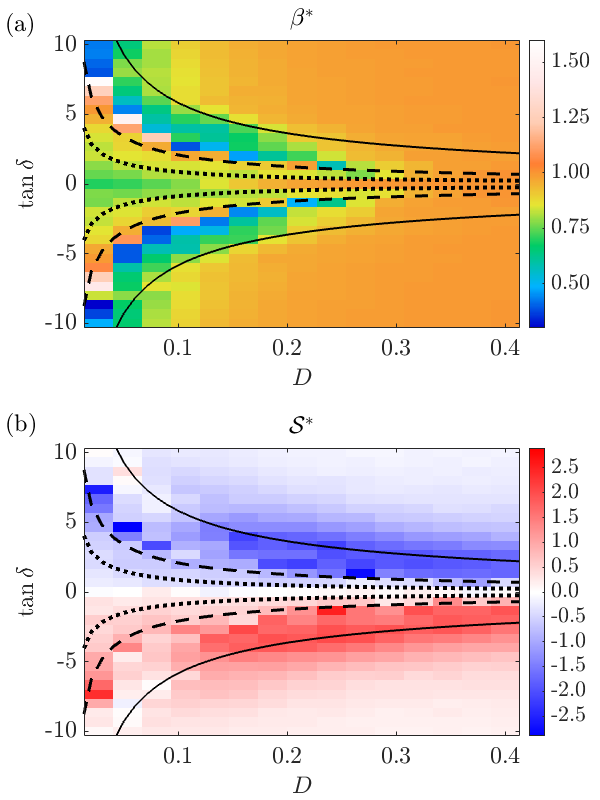}
\caption{{\sf \bf Dynamic phase diagrams for rings of $L=1000$ KS oscillators in the presence of columnar disorder.} (a) Estimate of growth exponent $\beta^*$ as a function of the non-oddity $\tan \delta$ and the noise strength $D$. (b) Estimate of skewness of fluctuations $\mathcal{S}^*$ as a function of the same two parameters.  The solid/dashed/dotted black lines show $g^* = 2/0.1/0.01$ level curves, cf.\ Eqs.\ \eqref{KPZcoupl} and \eqref{gstar}. See Sec.~\ref{subsec:PD} for additional definitions and expected behaviors. Results based on 500 realizations.}
\label{PDdyn_col}
\end{figure}

In Fig.~\ref{PDdyn_col} we show the estimate of the growth exponent $\beta^*$ [panel (a)] and of the skewness $\mathcal{S}^*$ [panel (b)] as functions of the control parameters, including the same level curves for $g^*$. For large $g^*$, in those regions that are far from the synchronization-desynchronization boundary in Fig.~\ref{PDstat_col}, we now find $\beta^* \approx \beta_\text{LG} = 1$ and modest values of $S^*$, as expected. But as the boundary from desynchronization is approached from outside, extremely large values of the skewness $\mathcal{S}^*$ are found, while $\beta^*$ is not yet far from $\beta_\text{LG} = 1$. This happens for large $D$ and moderate $\tan \delta$, but not for small $D$ and large $\tan \delta$, suggesting that different routes out of synchronization for the same value of the (approximate) KPZ coupling $g^*$ (\ref{gstar}) may not be equivalent under columnar disorder. The nature of this nonsynchronous dynamics with heavily asymmetric fluctuations for large $D$, which arise from the perturbation of the (synchronous) faceted profiles reported in Refs.~\cite{moroney,PRR1}, is addressed in Appendix \ref{appskewcol}.

In the presence of synchronization the situation observed in Fig.~\ref{PDdyn_col} is also considerably less clear than for time-dependent noise, which may be partly due to the non-universal corrections to columnar KPZ scaling mentioned in Sec.~\ref{subsec:rough}. For small $g^*$ we find values compatible with cEW universality (previously reported for $\delta = 0$ \cite{PRR1}  and other odd couplings \cite{PhysicaD}), namely $\beta^* \approx \beta_\text{cEW} = 3/4$ and $S^* \approx 0$. Yet for larger $g^*$ (see the region comprised between the two level curves for $g^* =0.01$ and $0.1$), there is a clear non-EW behavior. The values of $\beta^*$ there change gradually from values below $\beta_\text{cEW}$ to values above as $g^*$ increases, perhaps indicating non-universality of $\beta_\text{cKPZ}$ \cite{PRR1}. Phase slips, which also occur near the desynchronization boundary in this case, and also (trivially) over the nonsynchronous region, see Appendix \ref{appslips}, are also likely to play a role here.

\begin{figure}[t!]
\includegraphics[scale=0.42]{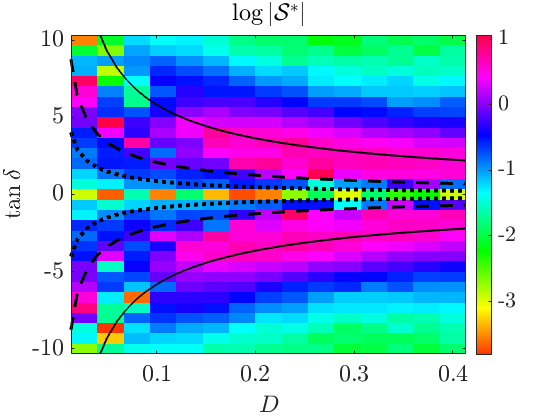}
\caption{{\sf \bf Skewness of fluctuations in logarithmic scale for rings of $L=1000$ KS oscillators in the presence of columnar disorder.} Here $\log|\mathcal{S}^*|$ is shown as a function of the non-oddity $\tan \delta$ and the noise strength $D$. Plot based on the same data for the skewness $\mathcal{S}^*$ included in Fig.~\ref{PDdyn_col} (b), with the same level curves for $g^*$.}
\label{logskew_col}
\end{figure}

Moreover, in Fig.~\ref{PDdyn_col} (b) for $g^*$ less than $0.1$ (dashed line) yet sufficiently larger than $0$, $\mathcal{S}^*$ takes on a value which appears to be roughly compatible with $\mathcal{S}_\text{GOE-TW} \approx 0.29$, as reported in Ref.~\cite{PRR1}, although this is also less clear than for the time-dependent noise case of Fig.~\ref{PDdyn_td} due to different range of inspected values, and could also be affected by non-universal features. To provide a better visualization, in Fig.~\ref{logskew_col} we show a modified version of Fig.~\ref{PDdyn_col} (b), where instead of $\mathcal{S}^*$ we display $\log|\mathcal{S}^*|$ as a function of the control parameters. Considering that $\log|\mathcal{S}_\text{GOE-TW}|\approx -1.24$, which in the color scheme appears as dark blue, we confirm that skewness values compatible with GOE-TW are to be found in the region for which $g^* < 0.1$ for nonodd coupling with $\delta\neq 0$.

 The reader may have noticed that the changes undergone by the phase diagrams in Figs.~\ref{PDstat_col} and \ref{PDdyn_col} under a sign reversal of the nonoddity, $\delta \to - \delta$, are analogous to those reported for time-dependent noise in the previous subsection. These phase diagrams correspond to rings of $L=1000$ oscillators, but other system sizes, including $L= 200$ and $500$, have also been inspected for columnar disorder, with no substantial qualitative differences being observed (not shown). To illustrate this point, in Appendix \ref{appsmall} we show results for a smaller ring containing $L=100$ oscillators, which look quite similar to those reported here for $L=1000$, except for a larger cEW-dominated region and even larger skewness $S^*$ around the boundary. Moreover, there is the difference that, for $\delta=0$, the noise strength $D$ at which synchronization is lost decreases with the system size, in agreement with the dependence of the critical coupling for synchronization dependence on $L$ rigorously established in Ref.~\cite{strogatz}. 

We find that the columnar KPZ behavior reported in Refs.~\cite{PRR1,PRR2,PhysicaD} is relatively delicate, as it lies between a crossover to columnar EW behavior (which becomes more prevalent for smaller system sizes) and the first signs of desynchronization appearing under the form of the above-mentioned phase slips, just as happened for time-dependent noise. On top of that, the non-universality issues associated with the columnar KPZ class make the unquestionable observation of such regime less certain, despite the fact that it has been convincingly observed for both phase \cite{PRR1} and some limit-cycle oscillators \cite{PRR2}, for parameter values that, thanks to the present study, we now know are not far from the desynchronization boundary.

\section{Summary and Conclusions}
 
We have investigated the nonequilibrium criticality of synchronization in one dimension, under time-dependent noise or columnar disorder, by an extensive numerical study of phase oscillators in a ring coupled through a KS coupling function. The results are condensed into phase diagrams that include static features (defining the region over which synchronization takes place) and critical-dynamics features (characterizing the power-law growth towards synchronization and the asymmetry of the fluctuations around it). Following recent work also aimed at elucidating the universal features of synchronization in one dimension, our phase diagrams are based on adaptations of well-established observables in the kinetic-roughening literature, with the phase field here playing the role of the height field in interfacial growth contexts. Two control parameters are considered, namely, the randomness strength (relative to the coupling strength) and the nonoddity of the coupling function, which has been shown to be relevant both for the existence of a coupling threshold for synchronization \cite{strogatz,ostborn} and for the emergence of nonequilibrium critical dynamics at large space-time scales \cite{PRR1,PRR2,PRE,PhysicaD}.

\begin{figure}[t!]
\includegraphics[scale=0.42]{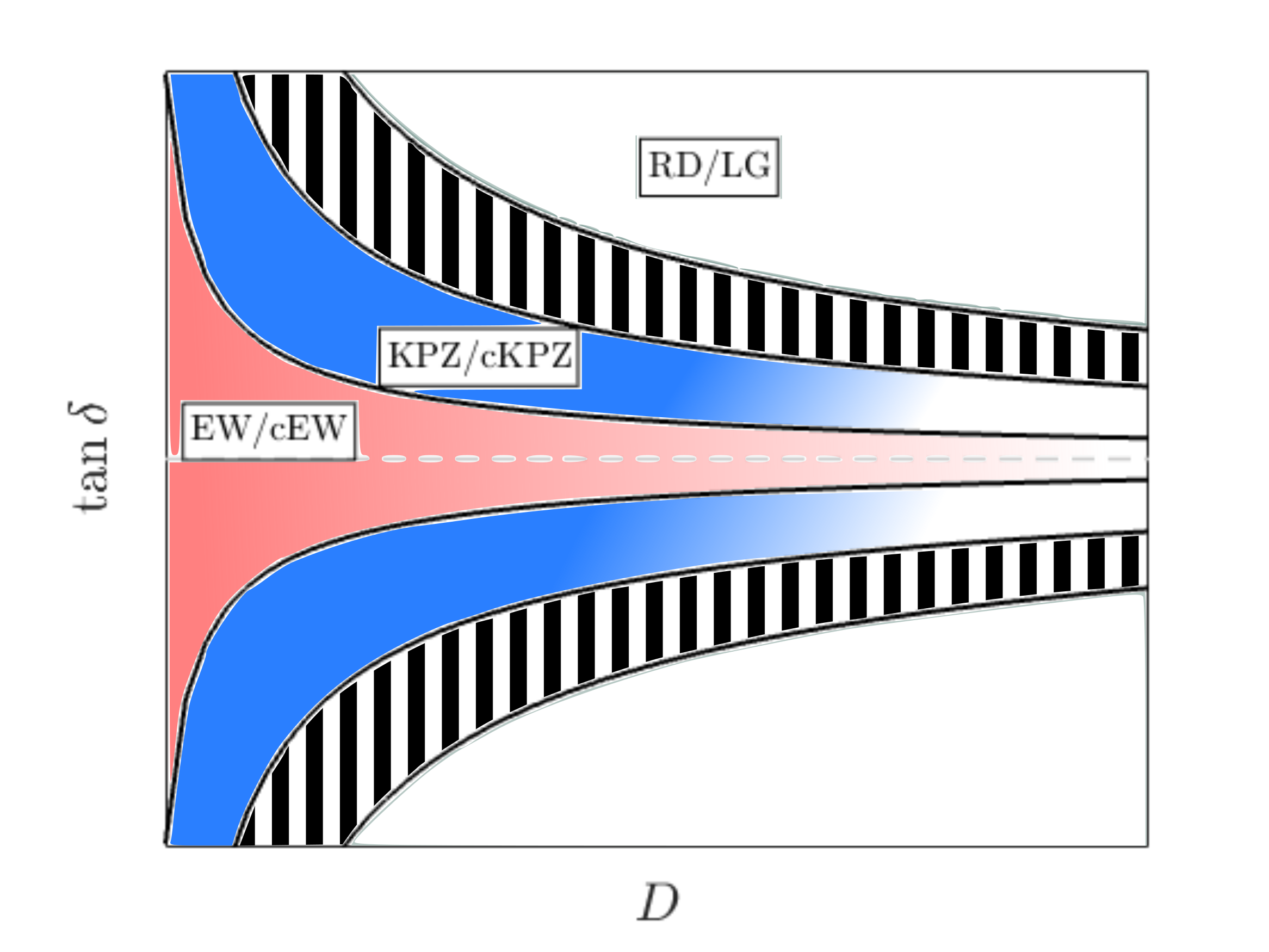}
\caption{{\sf \bf Schematic diagram showing the dynamical regimes at play in one-dimensional synchronization.} For small enough non-oddities $\tan \delta$ and/or randomness strength $D$ (red region), the behavior is compatible with EW (cEW) for synchronization under time-dependent noise (columnar disorder). As the parameters increase (blue region), we observe KPZ (cKPZ). Far into the nonsynchronous dynamics (white region) one finds RD (LG) behavior. In between the last two regimes there is an intermediate synchronization-desynchronization boundary region (thatched region), displaying a combination of synchronized behavior disrupted by very frequent phase slips and remnants of synchronization phenomonology in nonsynchronous dynamics, both of which defy any simple characterization. Within this region, EW (cWE) and KPZ (cKPZ) behaviors gradually disappear as $D$ or $\tan \delta$ are increased and synchronization is lost. The solid lines separating regions are well approximated by level curves of the effective KPZ coupling $g$, Eqs.\ \eqref{KPZcoupl} and \eqref{gstar}.}
\label{FigScheme}
\end{figure}

The general features observed can be best discussed by splitting the phase diagrams into three regions, a division that seems to be determined approximately by the strength of the KPZ coupling $g$ in the continuum limit, see Fig.~\ref{FigScheme} for a schematic representation. There is a clearcut region corresponding to nonsynchronous dynamics, and two for synchronous dynamics. The latter have these distinguishing features: well into the synchronization region (far from the synchronization boundary), a parameter-space region corresponding to small enough noise and nonoddity of the coupling, the dynamics is dominated by universal features compatible with the EW equation with the corresponding type of randomness; closer to the boundary (i.e.\! for larger values of the noise strength or the nonoddity), however, it is KPZ universality that dominates, though its observation requires a delicate choice of the control parameters, as well as sufficiently large system sizes. The nonsynchronous dynamics has the growth features of random deposition for time-dependent noise, while it displays a linear growth for columnar disorder (at least sufficiently far away from the desynchronization boundary). For time-dependent noise, some of these conclusions (the classification of distinct scaling behaviors in terms of $g$, specifically the occurrence of desynchronization for sufficiently large $g$ values) had been partly reached in Ref.\ \cite{lauter}.
 
It may come as a surprise that, in our phase diagrams, the KPZ behavior generically observed in previous publications, either for time-dependent noise \cite{PRE} or for columnar disorder \cite{PRR1,PRR2} (including a crossover from time-depedent KPZ to columnar KPZ when both forms of randomness are simultaneously present \cite{PhysicaD}), occupies a relatively narrow (albeit non-negligible, consistent with the claims on GSI) region in parameter space. While those previous references did report on the existence of a EW to KPZ crossover \cite{forrest}, which makes the KPZ behavior difficult to observe unless the nonoddity ---responsible for the KPZ nonlinearity in the continuum description--- is sufficiently strong and the system sufficiently large, no systematic study of such issues had been previously attempted. What we observe here is that, indeed, the observation of KPZ behavior may require setting the nonoddity to sufficiently large values (see also the equivalent discussion for phase-reduced dynamics of limit-cycle oscillators in \cite{PRR2}), especially for small randomness strengths, yet not so large that they bring about desynchronization. On top of that, there is the practical difficulty associated with the occurrence of phase slips very close to the desynchronization boundary, i.e.~the region where KPZ behavior is expected to be more easily observable. See the striped region in Fig.~\ref{FigScheme}, which includes both this synchronous motion affected by phase slips, but also the beginning of the nonsynchronous region, where remnants of synchronization further confound the picture. For a given strength of the randomness at play, the KPZ regime thus lies in a relatively narrow region between a EW crossover for too small nonoddities and the appearance of distorting phase slips for too large nonoddities, on the brink of desynchronization. That KPZ region is expected to become broader as the system size is increased, though, even further that we have been able to show here due to our computational limitations, since substantial (in terms of duration in time) EW-to-KPZ crossover is expected be confined to yet smaller nonoddities in larger systems. (The density of phase slips, depending on local fluctuations, should in principle be independent of the system size beyond certain value.)

To conclude, our present results, together with those previously reported in Refs.~{\cite{PRR1,PRR2,PRE,PhysicaD}}, imply that synchronization in one dimension is an instance of GSI. In those works every case studied with nonodd coupling (typically for systems of thousands of oscillators with a strong nonoddity) showed KPZ universality, and our present results confirm that such behavior is there, yet it may be elusive to observe unambiguously, unless the system is large and the parameters are adequately chosen.  Yet the present study is concerned with the practical observation of KPZ scaling in experiments and simulations of moderate sizes, and these are the main conclusions that we derive from it: (i) EW behavior may dominate also for $\delta \neq 0$ unless we are close enough to the desynchronization boundary or (possibly) for very large system sizes where the EW-to-KPZ crossover is left behind at the initial stages of the growth regime, (ii) even for the right sizes and parameter choices, practical difficulties for the study of KPZ scaling arise due to the appearance of phase slips very close to the desynchronization boundary. A practical course of action for the observation of GSI in synchronization (in numerical simulations or experiments), may thus be to first seek EW scaling in the data, using moderate nonoddities and sizes. And only then, by a judicious increase of the nonoddity or moving on to studying larger systems, start looking for the (pervasive, yet somewhat elusive in practice) KPZ regime. 

In retrospect, this seems to constitute yet another instance of the practical difficulties of observing KPZ universality. Indeed, and perhaps surprisingly from a current perspective, in spite of the theoretical expectations on the relevance of KPZ scaling, it took more than ten years to unambiguously identify this universality class in experiments \cite{takeuchi}. At the time that was due to competing physical effects (like, e.g., transport mechanisms of the required material to the growing surface) both in experimental and physical modeling contexts, and to difficulties in accessing sufficiently wide space and time scaling ranges, often due to crossover effects and/or dynamical instabilities \cite{cuerno04,Cuerno07} akin to those we have just characterized in this work.
  
 \section*{Acknowledgements}
We thank Gonzalo Contreras-Aso and Alejandro Vallejo Aparicio for insightful discussions and carefully reading the manuscript. This work has been partially supported by Ministerio de Ciencia e Innovaci\'on (Spain), by Agencia Estatal de Investigaci\'on (AEI, Spain, 10.13039/501100011033), and by European Regional Development Fund (ERDF, A way of making Europe) through Grants No.\ PID2021-123969NB-I00 and No.\ PID2021-128970OA-I00, and also through Grant No.\ PID2024-159024NB-C21, funded by MCIU/AEI/10.13039/501100011033/FEDER, EU. The support and computational resources of the Universidad Carlos III de Madrid (UC3M) C3 Cluster HPC, co-financed through action EQC2021-007184-P, funded by MICIU/AEI/10.13039/501100011033 and by the European Union NextGenerationEU/PRTR, is gratefully acknowledged.

\appendix

\makeatletter
\renewcommand\thesection{\Alph{section}}  % use plain letters A, B, C...
\makeatother

\section{Saturation times}
 \label{appsat}

The estimation of the saturation time $t^*$, as explained in Sec.~\ref{simsat}, is illustrated graphically in this appendix, and compared with theoretical estimates that are known to give reasonable results for the KPZ equation with time-dependent noise, and also for the EW equation with either time-dependent noise or columnar disorder. We will give evidence supporting the claim that the numerical criterion that we employ suffices for our purposes: by its application, we can identify the occurrence of saturation (synchronization) and the order of magnitude of the time at which it is reached.

Theoretical estimates for the KPZ equation with time-dependent noise arise from the following reasoning, which we adapt from Refs.~\cite{krug92,krug97}. Under the assumptions discussed in those references, based on dimensional considerations, the roughness grows as 
\begin{equation}
\mathcal{W}(t) = c_2^{1/2} A^{2/3} |\lambda|^{1/3} t^{1/3},
\label{hh}
\end{equation}
where $A = D^2/2\nu$, $\nu$, and $\lambda$ are the KPZ parameters in Eq.~(\ref{coarse}) and $c_2$ is a numerical constant. The growth exponent is then $\beta_{\rm KPZ}=1/3$, as in Table \ref{tab:exps}.  On the other hand, for large enough $L$ the saturation value of the roughness is 
\begin{equation}
\mathcal{W}_\text{sat} \approx (A L/12)^{1/2},
\label{Wsat}
\end{equation}
consistent with $\alpha_{\rm KPZ}=1/2$ as per Table \ref{tab:exps}. Equating (\ref{hh}) to the approximate saturation value (\ref{Wsat}) and solving for $t$ yields an estimate of the saturation time,
\begin{equation}
t_s^\text{KPZ} = \frac{L^{3/2}}{(12 c_2)^{3/2} |\lambda| A^{1/2}},
\label{tsKPZ}
\end{equation}
consistent with $z_{\rm KPZ}=3/2$, see Table \ref{tab:exps}. As for several models it has been numerically found that $c_2 \approx 0.4$ provides a good fit to the data, we will take Eq.~(\ref{tsKPZ}) with that choice for $c_2$ as our theoretical estimate of the saturation time for KPZ with time-dependent noise.

In the case of the EW equation with time-dependent noise, the saturation value is still given by Eq.\ (\ref{Wsat}) as, due to an accidental fluctuation-dissipation relation that only occurs for one-dimensional substrates, the 1D KPZ equation shares the stationary-state statistics of the 1D EW equation \cite{barabasi}. In the growth regime, however, the EW roughness grows as
\begin{equation}
\mathcal{W}(t) = \frac{A^{1/2}}{\pi^{1/4}} (2\nu t)^{1/4},
\end{equation}
for sufficiently large $L$ \cite{krug97}, implying $\beta_{\rm EW}=1/4$, see Table \ref{tab:exps}. By equating this expression to (\ref{Wsat}), the following estimate of the saturation time is obtained,
\begin{equation} 
t_s^\text{EW} = \frac{\pi L^2}{288 \nu},
\label{tsEW}
\end{equation}
leading to $z_{\rm EW}=2$, see Table \ref{tab:exps}.

Saturation time estimates in Eqs.~(\ref{tsKPZ}) and (\ref{tsEW}) are both given in terms of the  $\nu$, $\lambda$, and $D$ parameters. In order to use these formulae, we need to give some values to those parameters; we simply choose those that arise from the continuum approximation leading to Eq.~(\ref{coarse}) for the KS coupling in Eq.~(\ref{gamma}). Taking the lattice spacing to be $a=1$ and the coupling strength to be $K=1$, we obtain $\nu = \cos \delta$ and $\lambda = -2\sin \delta$, while we assume that $D$ is just the noise strength of the oscillators. Those parameters may well renormalize in a nontrivial fashion when going to larger space-time scales, yet, for lack of a detailed theory, we take them straight out from the (microscopic) oscillator model.

\begin{figure}[t!]
\includegraphics[scale=0.42]{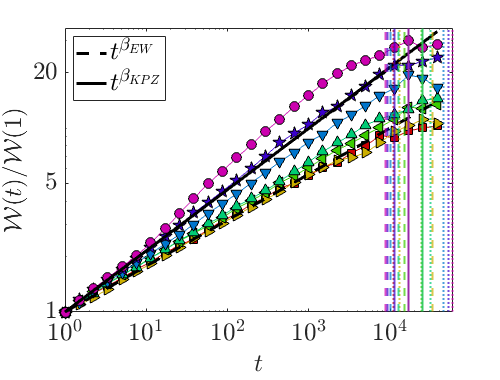}
\caption{{\sf \bf Saturation times in a ring of $L=1000$ oscilators with time-dependent noise.} Normalized roughness $\mathcal{W}(t)/\mathcal{W}(1)$ as a function of time $t$ for noise strength $D=0.1066$ and nonoddities $\tan \delta = 0$ (red squares), $2/3$ (yellow right triangles) up to $16/3$ (magenta circles) in steps of $2/3$, with larger $\tan \delta$ results consistenly above results for lower $\tan \delta$. Vertical lines %for value of $\tan \delta$ 
represent saturation times for the corresponding color of the symbols that encodes the same value of $\tan \delta$: numerical estimate based on the roughness data, $t^*$ (solid), KPZ theoretical estimate (dashed), see Eq.~(\ref{tsKPZ}),  and EW theoretical estimate (dotted), see Eq.~(\ref{tsEW}). Results based on 500 realizations.}
\label{Sattimes_td}
\end{figure}

In Fig.~\ref{Sattimes_td} we show the roughness $\mathcal{W}(t)$ for systems of $L=1000$ oscillators in the presence of time-dependent noise of strength $D=0.1066$, and several values of $\tan \delta$ (curves for larger $\tan \delta$ lying above curves for smaller $\tan \delta$). While the second and third largest $\tan \delta$ considered shows KPZ scaling, at least values for smaller $\tan \delta$ are still in the EW regime (while the largest non-oddity shows distortion due to phase slips). Even larger values of the nonoddity $\tan \delta$ (following the same spacing of $2/3$) are not included as for them the system does not saturate (see Appendix \ref{appslips}), while results for negative $\tan \delta$ are excluded as they are essentially indistinguishable from their positive counterparts, just as expected. For each choice of $\delta$ we display the numerical estimate of the saturation time $t^*$ (see Sec.~\ref{simsat}) as a solid vertical line, as well as the theoretical estimates in Eqs.~(\ref{tsKPZ}) and (\ref{tsEW}), represented as dashed and dotted vertical lines, respectively, all of them in the same color as the numerical data for the corresponding $\tan \delta$. Some lines appear to be absent due to overlaps. From this and similar plots for other system sizes and noise strengths, we conclude that the theoretical estimates are not too different from the numerically found $t^*$, all of them lying in a region where the roughness $\mathcal{W}(t)$ is reaching its plateau.

No theoretical estimates for the saturation time in the presence of columnar disorder are to be found in the literature, as far as we know. But in the case of the columnar EW equation, one can be found by noticing that the analytical solution for the structure factor \cite{purrello},
\begin{equation}
S(k,t) = \frac{2\pi D^2}{\nu^2 k^4}\left(1-e^{-\nu k^2 t}\right)^2,
\label{SkCEW}
\end{equation}
when integrated over all wavenumbers in the first Brillouin zone $[-\pi/a,\pi/a]$ yields the squared roughness, thus
\begin{align}
\mathcal{W}^2(t)  &= \sum_{k\neq 0} S(k,t) \approx 2 \int^{\frac{\pi}{a}}_{\frac{2\pi}{L}} \frac{d k}{2\pi} S(k,t)\nonumber \\ 
&= \frac{2 D^2}{\nu^2}\int^{\frac{\pi}{a}}_{\frac{2\pi}{L}}\frac{dk}{ k^4}\left(1-e^{-\nu k^2 t}\right)^2 ,
\label{W2CEW}
\end{align}
%\textcolor{red}{¿Está bien normalizado el $dk/2\pi$?}
using that there are $L$ reciprocal wavevectors $k_n = 2\pi n/L a$ for $n = 0,\pm 1,\cdots, \pm L/2$ and $L\gg 1$. %\textcolor{red}{Es correcto?}) yields the roughness squared, 
In the $t\to \infty$ limit, the exponential vanishes yielding the saturation value
\begin{equation}
\mathcal{W}^2_\text{sat} = \frac{2 D^2}{\nu^2} \int^{\frac{\pi}{a}}_{\frac{2\pi}{L}} \frac{d k}{k^4} = \frac{2 D^2}{\nu^2} \left[-\frac{1}{3 k^{3}} \right]^{\frac{\pi}{a}}_{\frac{2\pi}{L}} \approx \frac{2 D^2 L^3}{3 (2\pi)^3 \nu^2},
\label{W2CEWsat}
\end{equation}
%as $L^3 \gg a^3$.
consistent with $\alpha_{\rm cEW}=3/2$ as in Table \ref{tab:exps}. For the finite-time expression (\ref{W2CEW}), %as we are interested in intermediate times, for which the dominant wavenumbers satisfy $2\pi/L \ll k \ll \pi/a$ \textcolor{red}{(¿Se puede argumentar mejor? Si no, en Mathematica imagino que no hay problema en tomar los límites de integratción tal cual quedan al hacer el cambio de variable, ya que el integrando tiene una primitiva analítica con funciones de error.)}, 
$L\to \infty$ and $a\to 0$ can be safely taken in the integration limits; changing variables as $y = \nu k^2 t$ then yields
\begin{equation}
\mathcal{W}^2(t)  = \frac{2 D^2 t^{3/2}}{2 \nu^{1/2}} \int^{\infty}_{0}\frac{d y}{y^{5/2}} \left(1-e^{-y}\right)^2 \approx \frac{2 D^2 t^{3/2}}{\nu^{1/2}},
\label{W2CEW2}
\end{equation}
since the integral can be computed using the gamma function, yielding $\frac{8}{3} (\sqrt{2}-1)\sqrt{\pi} \approx 1.96$, in agreement with $\beta_{\rm cEW}=3/4$ as in Table \ref{tab:exps}. By equating this expression to Eq.~(\ref{W2CEWsat}) and solving for the time variable, we find an estimate of the saturation time,
\begin{equation}
t_s^\text{cEW} = \frac{L^2}{3^{2/3} (2\pi)^2 \nu}, 
\label{tsCEW}
\end{equation}
which is numerically close to the time-dependent EW esimate (\ref{tsEW}) and consistent with $z_{\rm cEW}=2$, see Table \ref{tab:exps}.

\begin{figure}[t!]
\includegraphics[scale=0.42]{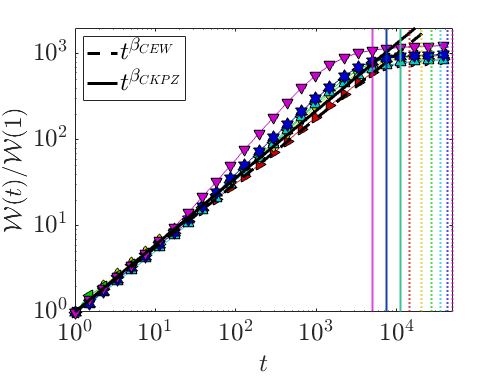}
\caption{{\sf \bf Saturation times in a ring of $L=1000$ oscilators with columnar disorder.} Normalized roughness $\mathcal{W}(t)/\mathcal{W}(1)$ as a function of time $t$ for noise strength $D=0.0533$ and nonoddities $\tan \delta = 2/3$ (red right triangles), $4/3$ (yellow diamonds) up to $12/3 = 4$ (magenta down triangles) in steps of $2/3$, with larger $\tan \delta$ results consistenly above results for lower $\tan \delta$.  Vertical lines %for value of $\tan \delta$ 
represent saturation times for the corresponding color of the symbols that encodes the same value of $\tan \delta$: numerical estimate based on the roughness data, $t^*$ (continuous) and columnar EW theoretical estimate (dotted), see Eq.~(\ref{tsCEW}). Results based on 500 realizations.}
\label{Sattimes_col}
\end{figure}

In Fig.~\ref{Sattimes_col} we show the roughness $\mathcal{W}(t)$ for systems of $L=1000$ oscillators in the presence of columnar disorder of strength $D=0.0533$ for several values of the nonoddity $\tan \delta$. % = 0$ (red), $2/3$ (yellow), $4/3$ (green), $2$ (light blue), $8/3$ (dark blue) and $10/3$ (magenta). 
The largest one is affected by phase slips (see Appendix \ref{appslips}) that distort the power-law behavior, while the similarity between $\beta^\text{cEW}$ and $\beta^\text{KPZ}$ (see Table \ref{tab:exps}) makes it hard to distinguish the behavior for the other values. The reasons for restricting ourselves to this set of $\tan \delta$ values is the same as discussed above in the time-dependent-noise setting. For columnar disorder, we only have theoretical estimates of the saturation time for the columnar EW equation, Eq.~(\ref{tsCEW}), which appear as vertical dotted lines, and not for the columnar KPZ equation, while the solid lines again represent the numerical estimates $t^*$. Also here some lines appear to be absent due to overlaps. The theoretical estimate $t_s^\text{cEW}$ seems to be too conservative in this case, as it is well into the saturation plateau, especially for large nonoddities (where probably a good columnar KPZ estimate would work better). As for the numerical method to estimate $t^*$, it seems to provide reasonable results in this and similar plots for other disorder strengths and sizes (not shown).

\section{Smaller systems}
\label{appsmall}

\begin{figure}[t!]
\includegraphics[scale=0.42]{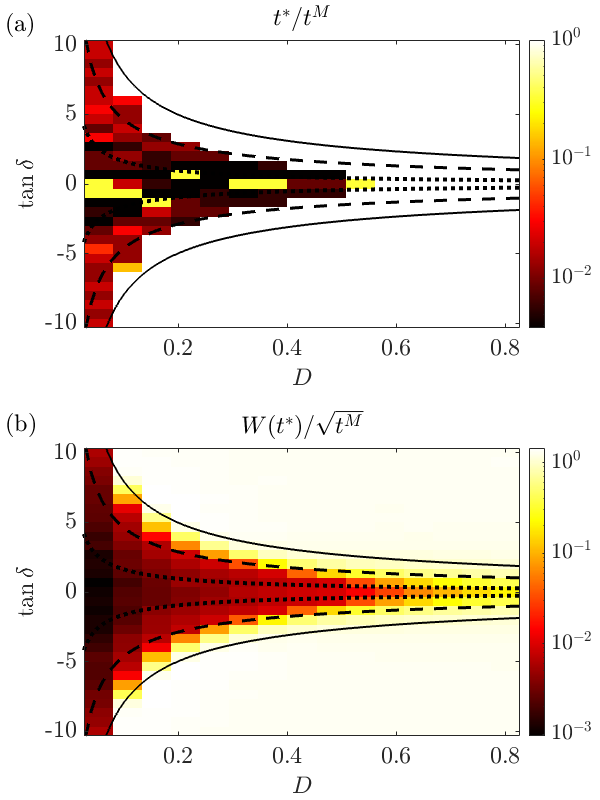}
\caption{{\sf \bf Static phase diagrams for rings of $L=100$ KS oscillators in the presence of time-dependent noise.} (a) Normalized estimate of the saturation time $t^*/t^M$ as a function of the non-oddity $\tan \delta$ and the noise strength $D$. (b) Normalized roughness at saturation $\mathcal{W}(t^*)/\sqrt{t^M}$ as a function of the same two parameters. The solid/dashed/dotted black lines show $g^* = 5/1/0.05$ level curves, cf.\ Eq.\ \eqref{gstar}. See Sec.~\ref{subsec:PD} for additional definitions and expected behaviors. Results based on 5000 realizations.}
\label{PDstat_td_100}
\end{figure}

\begin{figure}[t!]
\includegraphics[scale=0.42]{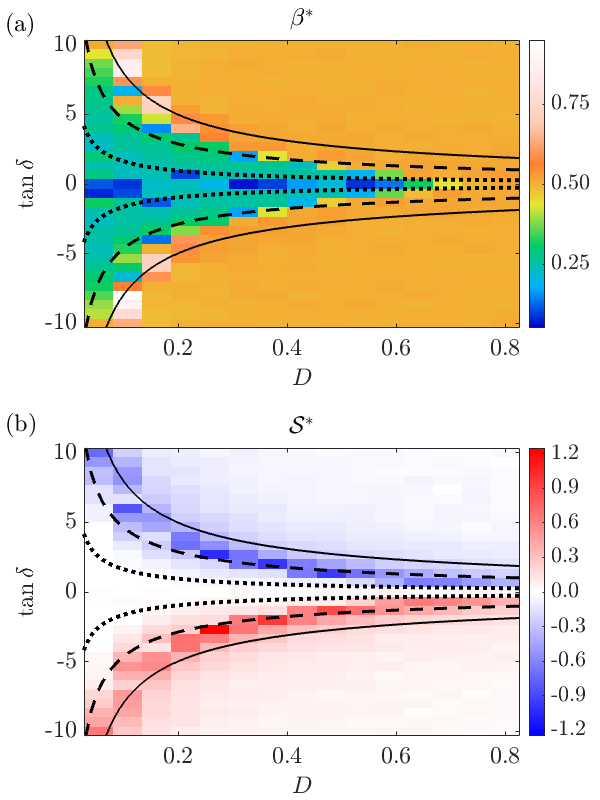}
\caption{{\sf \bf Dynamic phase diagrams for rings of $L=100$ KS oscillators in the presence of time-dependent noise.} (a) Estimate of growth exponent $\beta^*$ as a function of the non-oddity $\tan \delta$ and the noise strength $D$. (b) Estimate of skewness of fluctuations $\mathcal{S}^*$ as a function of the same two parameters. The solid/dashed/dotted black lines show $g^* = 5/1/0.05$ level curves, cf.\ Eq.\ \eqref{gstar}. See Sec.~\ref{subsec:PD} for additional definitions and expected behaviors. Results based on 5000 realizations.}
\label{PDdyn_td_100}
\end{figure}

Time-dependent-noise results analogous to those reported in the phase diagrams in Figs.~\ref{PDstat_td} and \ref{PDdyn_td} for systems of $L=1000$ oscillators, but for smaller systems of $L=100$ oscillators are displayed in Fig.~\ref{PDstat_td_100} and \ref{PDdyn_td_100}. Despite the considerable change in system size, we observe the same qualitative features. The static phase diagrams are very similar, with the somewhat trivial difference that the normalized saturation times $t^*/t^M$ in Fig.~\ref{PDstat_td_100} (a) are lower than in Fig.~\ref{PDstat_td} (a). The reason for this is simply computational: for rings of $L=100$ oscillators (smaller systems to numerically integrate, which moreover saturate earlier, given $t_\text{sat} = L^z$) we could take a much more conservative maximum simulation time $t^M$ than for $L=1000$.

The most conspicuous difference is in the growth results in Fig.~\ref{PDdyn_td_100}, particularly in the region close to desynchronization, which is less well defined and harder to ascribe to KPZ universality. Results for the growth exponent $\beta^*$ in Fig.~\ref{PDdyn_td_100} (a) are noisier, and those where $\beta^* \approx \beta_\text{KPZ} = 1/3$ occupy a smaller region in parameter space than in Fig.~\ref{PDdyn_td} (a). This is compatible with the existence of a larger region dominated by the crossover to EW universality (characterized by $\beta^* \approx \beta_\text{EW} = 1/4$ and $\mathcal{S}^* \approx 0$) in these smaller systems. The skewness $\mathcal{S}^*$ values around that region in Fig.~\ref{PDdyn_td_100} (b) also appear to be somewhat smaller than those in Fig.~\ref{PDdyn_td} (b).

\begin{figure}[t!]
\includegraphics[scale=0.42]{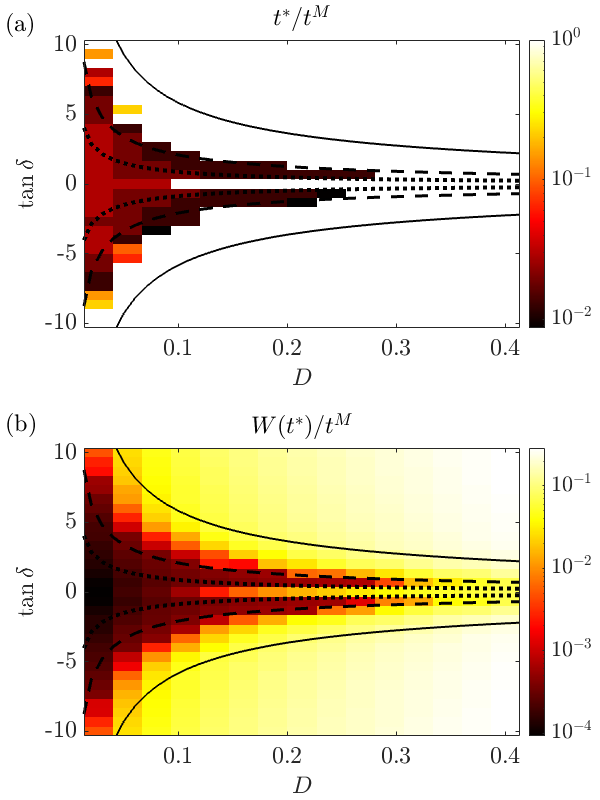}
\caption{{\sf \bf Static phase diagrams for rings of $L=100$ KS oscillators in the presence of columnar disorder.} (a) Normalized estimate of the saturation time $t^*/t^M$ as a function of the non-oddity $\tan \delta$ and the noise strength $D$. (b) Normalized roughness at saturation $\mathcal{W}(t^*)/\sqrt{t^M}$ as a function of the same two parameters. The solid/dashed/dotted black lines show $g^* = 2/0.1/0.01$ level curves, cf.\ Eq.\ \eqref{gstar}. See Sec.~\ref{subsec:PD} for additional definitions and expected behaviors. Results based on 5000 realizations.}
\label{PDstat_col_100}
\end{figure}

\begin{figure}[t!]
\includegraphics[scale=0.42]{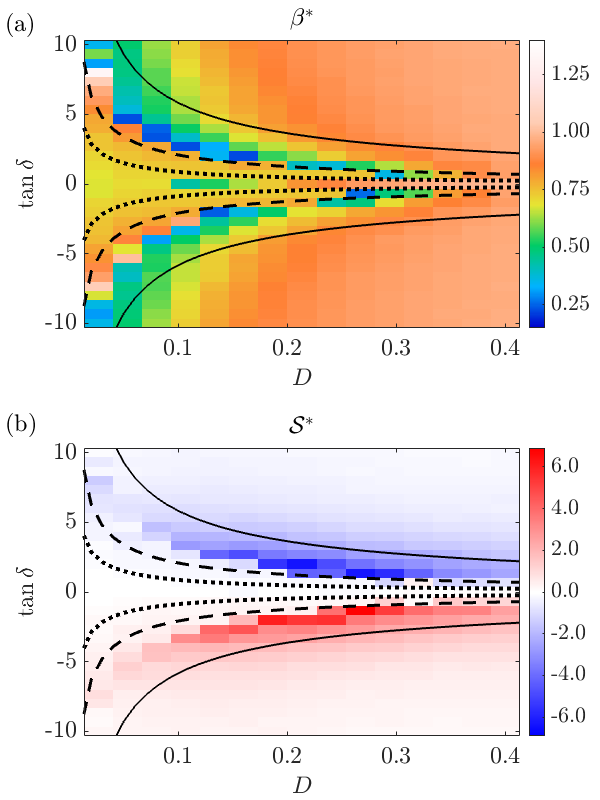}
\caption{{\sf \bf Dynamic phase diagrams for rings of $L=100$ KS oscillators in the presence of columnar disorder.} (a) Estimate of growth exponent $\beta^*$ as a function of the non-oddity $\tan \delta$ and the noise strength $D$. (b) Estimate of skewness of fluctuations $\mathcal{S}^*$ as a function of the same two parameters. The solid/dashed/dotted black lines show $g^* = 2/0.1/0.01$ level curves, cf.\ Eq.\ \eqref{gstar}. See Sec.~\ref{subsec:PD} for additional definitions and expected behaviors. Results based on 5000 realizations.}
\label{PDdyn_col_100}
\end{figure}

Columnar-disorder results analogous to those reported in the phase diagrams in Figs.~\ref{PDstat_col} and \ref{PDdyn_col} for systems of $L=1000$ oscillators, but for smaller systems of $L=100$ oscillators are displayed in Fig.~\ref{PDstat_col_100} and \ref{PDdyn_col_100}. In the static phase diagram of Fig.~\ref{PDstat_col_100} (a), for $\tan \delta = 0$ ($\delta = 0$) the system desynchronizes for a larger noise strength $D$ that in  Fig.~\ref{PDstat_col} (a). This a feature of synchronization in the presence of columnar disorder, reflecting the fact that larger systems require stronger couplings, or equivalently weaker noise, in order to reach synchronization \cite{PRR1,strogatz}. Besides this, there is also here the (trivial) effect of the more conservative choice of $t^M$ for the smaller systems, leading to the darker colors observed for $t^*/t^M$ in Fig.~\ref{PDstat_col_100} (a) relative to Fig.~\ref{PDstat_col} (a).

Regarding the growth features, in Fig.~\ref{PDdyn_col_100} (a) the region displaying columnar EW growth as given by $\beta^*$ appears to be larger than in Fig.~\ref{PDdyn_col} (a), which is compatible with enhanced crossover effects for smaller systems. In the case of the skewness $S^*$, Fig.~\ref{PDdyn_col_100} (a) displays even larger extreme values for $L=100$ oscillators than shown for $L=1000$, in the dynamical regime where fluctuations around the linear growth are heavily asymmetric. That phenomenology is partially elucidated in Appendix \ref{appskewcol}.

 \section{Phase slips}
  \label{appslips}

We have mentioned that one of the difficulties associated with the observation of KPZ universality, for either type of randomness, is that it happens when the nonoddity $\tan \delta$ is large enough in absolute value (precisely how large depends on the randomness strength $D$), or when $D$ is large enough (precisely how large depends on $\tan \delta$). In a coarse grained description, that seems to imply that the KPZ nonlinearity in Eq.~(\ref{coarse}) is sufficiently prominent so that the crossover from EW to KPZ behavior \cite{forrest} is observed before saturation sets in, which requires a strong coupling $g^*$ (\ref{gstar}). But increasing the KPZ coupling $g^*$ also makes the phase profile less stable, which may eventually lead to desynchronization. See Figs.~\ref{PDdyn_td} and \ref{PDdyn_col}, where KPZ (or columnar KPZ) behavior appears rather close to the boundary between synchronous and nonsynchronous dynamics. 

For a large value of $g^*$, just before desynchronization proper is reached, the phase profile given by $\{\phi_i(t)\}_{i=1}^L$ develops some ``discontinuities'' (i.e.~sudden jumps between phases lying at neighboring sites) which are known in the literature as phase slips \cite{pikovsky}. In fact, as the coupling function $\Gamma(\Delta \phi)$ is $2\pi$-periodic, any relatively strong perturbation on an individual oscillator that makes its phase rapidly perform one or more full rotations (either clockwise or counterclockwise) introduces a phase increase of $2\pi k$ for $k\in \mathbb{Z}$ with respect to its neighbors that essentially leaves the interactions unchanged. As there is no restoring mechanism for such slips, and given the small probability of a second such event that perfectly compensates for the first one, those perturbations are irreversibly  retained in the phase profile. They cause a distortion of the scaling behavior as given by, e.g., the roughness $\mathcal{W}(t)$, the likes of which do not typically arise in the physics of kinetic roughening. In fact, couplings between neighboring sites in that context are typically increasing functions of the local slopes, and certainly not periodic functions.

\begin{figure}[t!]
\includegraphics[scale=0.42]{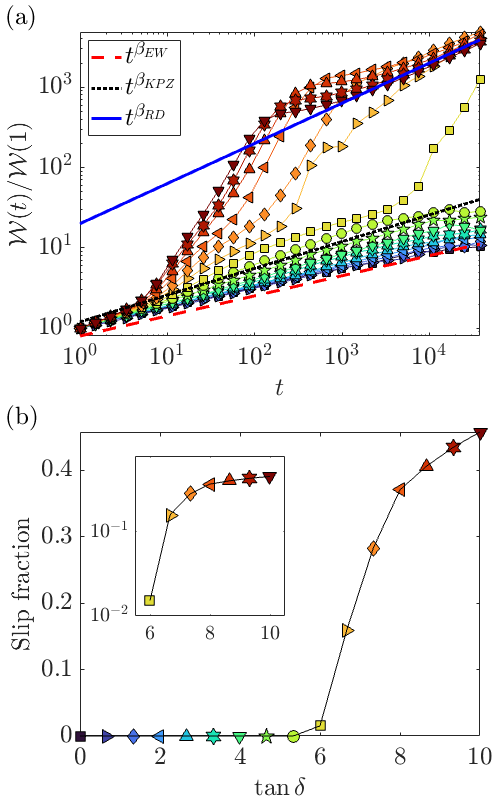}
\caption{{\sf \bf Roughness and phase slips in a ring of $L=1000$ oscillators in the presence of time-dependent noise} (a). Normalized roughness $\mathcal{W}(t)/\mathcal{W}(1)$ as a function of $t$ for noise strength $D=0.1066$ and values of the nonoddity $\tan \delta \in \{0,2/3,4/3,2,...,28/3, 10\}$ [see panel (b), where the value of $\tan \delta$ for each symbol and color can be read from the abcissae]. The straight lines represent the scaling corresponding to EW, KPZ, and LG universalities as indicated in the legend. (b) Slip fraction for the same values of $\tan \delta$. (Inset) Same plot in logarithmic scale (with zero values excluded). Results based on $500$ realizations.}
\label{Slips_td}
\end{figure}

To illustrate this issue, in Fig.~\ref{Slips_td} (a) we show the roughness $\mathcal{W}(t)$ for systems of $L=1000$ oscillators in the presence of time-dependent noise of strength $D=0.1066$. Lines displaying $t^\beta$ scaling for $\beta = \beta_\text{EW} = 1/4$, $\beta_\text{KPZ} = 1/3$, and $\beta_\text{RD} = 1/2$ are also displayed. Different colors and symbols are associated with different values of the nonoddity. %$\tan \delta \in \{0,2/3,4/3,2,...,28/3, 10\}$ [see panel (b), where their values can be read from the abscissas]
We limit the discussion to positive $\tan \delta$, as data points for negative values of this parameter are equivalent, according to the symmetries under sign reversal of $\delta$ discussed in the main text. For the smallest $\tan \delta$, leading to saturation of the roughness, these results were already displayed in Fig.~\ref{Sattimes_td}. We find that for low $\tan \delta$ growth appears to follow the EW exponent, and for large enough $\tan \delta$ it does follow the RD exponent, in agreement with the discussion of Fig.~\ref{PDdyn_td}. In between, there appears to be KPZ scaling, preceded by some values displaying intermediate (crossover) behavior between EW and KPZ.

In Fig.~\ref{Slips_td} (b), the fraction of phase slips for those same values of $\tan \delta$ is reported using the same color coding as in panel (a). Such a slip fraction is computed as the number of phase slips observed by monitoring the phase profiles for phase increments $|\phi_{i+1}(t) - \phi_i(t)|> 7\pi/4$ across time in different realizations and finally normalizing by the product of the size $L$, the number of time points, and the number of realizations. As such, it takes values in $[0,1]$ with $0$ corresponding to the absence of phase slips and $1$ to the extreme case in which for every time point of each realization one sees a difference between neighboring phases exceeding the above threshold for every single oscillator. For $\tan \delta = 6$ we start seeing phase slips in the system, in fact saturation is not achieved for long times [see panel (a)]. There are other choices of $D$ for which the smallest $\tan \delta$ displaying phase slips still leads to saturation, and the roughness $\mathcal{W}(t)$ appears to be distorted as a result, as well as some cases where the smallest $\tan \delta$ affected by slips already leads to a clear RD behavior asymptotically (not shown). In the dynamical phase diagram in Fig.~\ref{PDdyn_td} (a) such phase slips observed for $D=0.1066$  and $\tan \delta = 6$ lead to a value $\beta^* \approx 0.38$, in between $\beta_\text{KPZ}$ and $\beta_\text{RD}$.

For even larger values of $\tan \delta$ we do see the slip fraction increase, Fig.~\ref{Slips_td} (b), as corresponds to the unsynchronous dynamics observed in those cases, see panel (a). For robust nonsynchronous dynamics, phase differences are expected to increase with time anyway, and calling them phase slips is an abuse of language due to our operational definition of such events, based on the simple criterion $|\phi_{i+1}(t) - \phi_i(t)|> 7\pi/4$. More interestingly, in such cases RD growth for sufficiently long times is preceded by an initial transient where the dynamics follows KPZ scaling, see Fig.~\ref{Slips_td} (a). This type of behavior en route out of synchronization has been previously studied in Ref.~\cite{lauter}, and it is the reason why for such large values of $\tan \delta$ in Fig.~\ref{PDdyn_td} (a) the exponent $\beta^*$ is observed to oscillate wildly between $0.2$ and values close to $1.0$.

\begin{figure}[t!]
\includegraphics[scale=0.42]{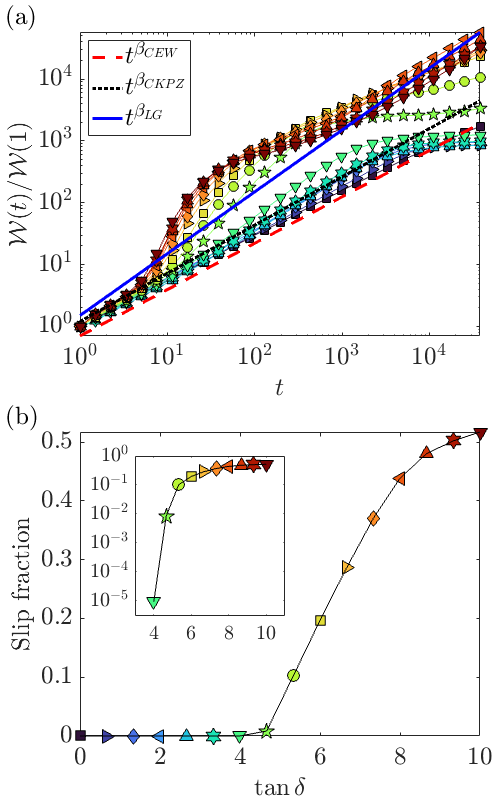}
\caption{{\sf \bf Roughness and phase slips in a ring of $L=1000$ oscillators in the presence of columnar disorder} (a)  Normalized roughness $\mathcal{W}(t)/\mathcal{W}(1)$ as a function of $t$ for disorder strength $D=0.0667$ and values of the nonoddity $\tan \delta \in \{0,2/3,4/3,2,...,28/3, 10\}$ [see panel (b), where the value of $\tan \delta$ for each symbol and color can be read from the abcissae]. The straight lines represent the scaling corresponding to EW, KPZ, and LG universalities as indicated in the legend.  (b) Slip fraction for the same values of $\tan \delta$. (Inset) Same plot in logarithmic scale (with zero values excluded). Results based on $500$ simulations.}
\label{Slips_col}
\end{figure}

Analogous results are reported in Fig.~\ref{Slips_col} for systems of $L=1000$ oscillators in the presence of columnar disorder. In that case we choose $D=0.0533$ for the disorder strength as it appears qualitatively similar to the time-dependent-noise strength of $D=0.1066$ in the phase diagrams (see Figs.~\ref{PDdyn_td} and \ref{PDdyn_col}). For small $\tan \delta$, these results were already displayed in Fig.~\ref{Sattimes_col}. With our numerical accuracy, it is very hard to distinguish between $\beta_\text{cEW}$ and $\beta_\text{cKPZ}$ growth, but in most cases the distintiction between such growth (which ends up in saturation) and $\beta_\text{LG}$ growth (corresponding to desynchronization) is clear. For $\tan \delta = 4$ we start seeing phase slips in the system, and the roughness $\mathcal{W}(t)$ appears to be distorted as a result, yet saturation is achieved (unlike what happens for larger $\tan \delta$). In the dynamical phase diagram in Fig.~\ref{PDdyn_col} this leads to the unusually large value $\beta^* \approx 1.04$ for $D=0.0533$ and $\tan \delta = 4$, even larger than $\beta_\text{LG}$.

\section{Highly asymmetric nonsynchronous dynamics under columnar disorder}
\label{appskewcol}

In the discussion of our columnar-disorder results we found an intriguing asymmetry in the fluctuations around the average growth for nonsynchronous dynamics. This is illustrated in Fig.~\ref{PDdyn_col} (b) for systems of $L=1000$ oscillators, where a very high skewness $S^*$, much higher than corresponds to a TW PDF, is observed on the nonsynchronous side of the synchronization-desynchronization boundary, especially for large noise strengths $D$. Even more extreme values are observed in Fig.~\ref{PDdyn_col_100} (b) for systems of $L=100$ oscillators around the same region in $(D,\tan \delta)$-space. In both cases, the corresponding values of the growth exponent $\beta^*$, Figs.~\ref{PDdyn_col} (a) and \ref{PDdyn_col_100} (a), appear to be somewhat below the linear growth exponent $\beta_\text{LG} = 1$, so the linear growth regime is not fully developed yet. In this appendix we offer some insight into these facts, leaving a deeper analysis for future work on routes out of desynchronization.

\begin{figure}[t!]
\includegraphics[scale=0.42]{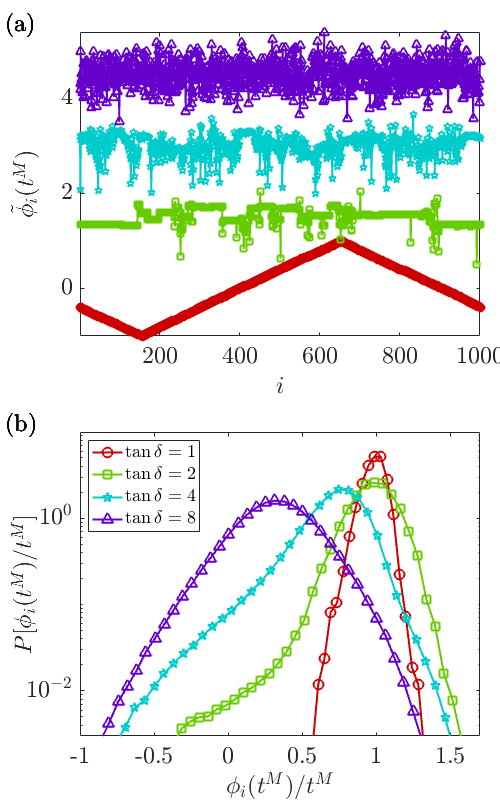}
\caption{{\sf \bf Phase profiles for long times $t= t^M = 10000$ in a system of $L=1000$ oscillators under columnar disorder of strength $D=0.3$.} (a)  Normalized phases (see text for definition) across space for $\tan \delta = 1,2,4$, and $8$ [see legend in panel (b)].  (b) Histograms of approximate effective frequencies $\phi_i(t^M)/t_M$ across $1000$ realizations of the columnar disorder for the same values of $\tan \delta$.}
\label{Skewcol}
\end{figure}

In Fig.~\ref{Skewcol} (a) we display the phase profiles for systems of $L=1000$ oscillators under columnar disorder of strength $D=0.3$, which is clearly affected by the high skewness mentioned above in Figs.~\ref{PDdyn_col} and \ref{PDdyn_col_100}, and nonoddities $\tan \delta = 1,2,4$, and $8$. Specifically, we display the normalized phases of a representative realization at the maximum simulation time considered $t^M = 10000$, defined as $\tilde \phi_i(t^M) = (\phi_i(t^M) - \overline{\phi_i(t^M)})/\max(|\phi_i(t^M)|)$, where the overbar denotes spatial average again, followed by a vertical displacement introduced for visilibity. As we are interested in the qualitative features of the morphologies, neither the average height nor the width (roughness) are important here. For $\tan \delta = 1$, the profile has the facets known to arise in synchronization under columnar disorder \cite{moroney,PRR1,PRR2}, which is very similar to the faceted behavior of the columnar KPZ equation \cite{szendro}. It turns out that synchronization under such form of quenched disorder develops as a coarsening process, in which nearby oscillators become locked to a common effective frequency, forming triangular (`faceted') arrangements that move in unison. Saturation is achieved when the largest (which is also the fastest) facet absorbs all oscillators, as shown in Fig.~\ref{Skewcol} (a) for $\tan \delta = 1$.

What we observe in Fig.~\ref{PDdyn_col} (a) for larger $\tan \delta$ is the result of facets that are not able to merge, due to the destabilizing influence of noise and the KPZ lateral growth mechanism. This is illustrated in the representative phase profile displayed in Fig.~\ref{Skewcol} (a) for $\tan \delta = 2$ or larger, showing much more irregular behavior with large phase jumps. Particularly for $\tan \delta =2$ or values closer to $\tan \delta = 1$ (not shown) the facets are shown to be disrupted by big jumps. As some of these oscillators evolve at different effective frequencies, the latter are not just the phase slips irreversibly created in the profile due to the presence of fluctuations that we discussed in the previous appendix.

To provide a more dynamical picture, in Fig.~\ref{PDdyn_col} (b) we represent the histogram of $\phi_i(t^M)/t^M$ for the same system based on $1000$ realizations of the columnar disorder. As $t^M$ is quite large, this not only corresponds to the phase profiles for long times normalized by a linear growth, but also can be interpreted as an estimate of the effective frequencies $\omega_\text{eff}$ in Eq.~(\ref{omegaeff}). We find that, indeed, for $\tan \delta = 1$, where the system synchronizes, the effective frequencies have a narrow distribution of values around the peak, while for $\tan \delta = 2$ and $4$ the distribution is highly skewed, which must correspond to the sudden jumps connecting the seemingly correlated segments in Fig.~\ref{PDdyn_col} (a). As expected from Fig.~\ref{PDdyn_col} (b), for even larger $\tan \delta$, the skewness of the distribution starts diminishing, leading to a more symmetric profile. Future work will be required to fully understand the implications of this unexpected dynamical behavior, which is peculiar to the oscillators under columnar disorder.

\end{document}